\documentclass{LMCS}

\def\doi{8 (3:11) 2012}
\lmcsheading%
{\doi}
{1--30}
{}
{}
{Feb.~12, 2011}
{Aug.~13, 2012}
{}

\usepackage{enumerate}
\usepackage{hyperref}
\usepackage{gastex}

\theoremstyle{plain}

\def\proofof#1{\rm \trivlist \item[\hskip \labelsep{\it
Proof of #1.}]}
\def\qedo{\popQED\endtrivlist}

\newcommand{\abs}[1]{\left|\mathinner{#1}\right|}
\newcommand{\Alpha}{\mathsf{alph}}

\newcommand{\mediumSize}[1]{\fontsize{9pt}{12pt}\selectfont #1\normalsize}
\newcommand{\mediumFont}[1]{\normalfont\mediumSize{#1}}
\newcommand{\malcev}%
  {\mathop{\text{\normalsize{\raisebox{0.1mm}{\textcircled{\mediumFont{M}}}}}}}
\font\petite=cmmi10 at 8pt
\def\malcev{\mathbin{\hbox{$\bigcirc$\rlap{\kern-9pt\raise0,75pt\hbox{\petite m}}}}}

\newcommand{\varietyFont}[1]{\mathrm{\mathbf{#1}}}
\newcommand{\DA}{\varietyFont{D\hspace{-1pt}A}}

\newcommand{\Jone}{\varietyFont{J_1}}

\newcommand{\Ap}{\varietyFont{A}}

\let\J\Jtrivial
\newcommand{\RR}{\varietyFont{R}}
\newcommand{\LL}{\varietyFont{L}}

\newcommand{\X}{\varietyFont{X}}
\let\V\Vara
\let\W\Varb
\newcommand{\VarFO}{\varietyFont{FO}}
\newcommand{\VarTL}{\varietyFont{TL}}
\newcommand{\K}{\varietyFont{K}}
\newcommand{\D}{\varietyFont{D}}

\newcommand{\LangV}{\mathcal{V}}  

\newcommand{\LangFO}{\mathcal{FO}}
\newcommand{\LangTL}{\mathcal{T\!L}}
\newcommand{\underLangTL}{\underline{\mathcal{T\!L}}}
\def\calJ{\mathcal{J}}
\def\calL{\mathcal{L}}
\def\calR{\mathcal{R}}

\newcommand{\greenFont}[1]{\mathcal{#1}}
\newcommand{\gR}{\mathrel{\greenFont{R}}}
\newcommand{\gL}{\mathrel{\greenFont{L}}}
\newcommand{\gJ}{\mathrel{\greenFont{J}}}

\newcommand{\FO}{\mathrm{\mathsf{FO}}} 

\newcommand{\TL}{\mathsf{TL}}  
\newcommand{\PTL}{\mathsf{PTL}}  

\newcommand{\XX}{\mathrm{\mathsf{X}}}
\newcommand{\YY}{\mathrm{\mathsf{Y}}}
\newcommand{\ZZ}{\mathrm{\mathsf{Z}}}

\newcommand{\TRUE}{\mathbf{\top}}

\newcommand{\RIGHT}{\mathrel{\triangleright}}
\newcommand{\LEFT}{\mathrel{\triangleleft}}

\def\inv{^{-1}}
\let\phi\varphi
\let\epsilon\varepsilon
\def\ord{\mathrm{ord}}

\def\underR{\underline{R}}
\def\underTL{\underline{\TL}}
\def\underVarTL{\underline{\VarTL}}

\def\Box{\hbox{\rlap{$\sqcap$}$\sqcup$}}
\def\block{\mathbin{\Box}}

\begin{document}

\title[Logical hierarchies within FO$^2$-definable languages]{On logical hierarchies within FO$^2$-definable languages\rsuper*}

\author[M.~Kufleitner]{Manfred Kufleitner\rsuper a}	
\address{{\lsuper a}University of Stuttgart, Germany}	
\email{kuf\-leitner@fmi.uni-stuttgart.de}  
\thanks{{\lsuper a}The first author was supported by the German Research Foundation (DFG) under grant \mbox{DI 435/5-1}}	

\author[P.~Weil]{Pascal Weil\rsuper b}	
\address{{\lsuper b}{Univ. Bordeaux, LaBRI, UMR 5800, F-33400 Talence, France\newline CNRS, LaBRI, UMR 5800, F-33400 Talence, France}}	
\email{pascal.weil@labri.fr}  
\thanks{{\lsuper b}The second author was supported by
    the grant ANR 2010 BLAN 0202 01 FREC}	


\titlecomment{{\lsuper*}An extended abstract of this paper has been published in the proceedings of MFCS 2009.}


\begin{abstract}
We consider the class of languages defined in the 2-variable fragment of the first-order logic of the linear order. Many interesting characterizations of this class are known, as well as the fact that restricting the number of quantifier alternations yields an infinite hierarchy whose levels are varieties of languages (and hence admit an algebraic characterization). Using this algebraic approach, we show that the quantifier alternation hierarchy inside $\FO^2[<]$ is decidable within one unit.  For this purpose, we relate each level of the hierarchy with decidable varieties of languages, which can be defined in terms of iterated deterministic and co-deterministic products.  A crucial notion in this process is that of condensed rankers, a refinement of the rankers of Weis and Immerman and the turtle languages of Schwentick, Th\'erien and Vollmer.
\end{abstract}

\maketitle

Many important properties of systems are modeled by finite automata. Frequently, the formal languages induced by these systems are definable in first-order logic. Our understanding of its expressive power is of direct relevance for a number of application fields, such as verification.

The first-order logic we are interested in, in this paper, is the first-order logic of the linear order, written $\FO[<]$, interpreted on finite words. It is well-known that the languages that are definable in this logic are exactly the star-free languages, or equivalently the regular languages whose syntactic monoid is aperiodic (that is: satisfies an identity of the form $x^{n+1} = x^n$ for some integer $n$) \cite{Schutzenberger1965sf,McNaughtonPapert1971book} (see also 
\cite{DiekertGastin2008siwt,Pin1986book,Straubing1994book,StraubingWeil2011Indianbk}); and that deciding whether a finite automaton accepts such a language is \textsc{PSPACE}-complete \cite{ChoHuynh1991tcs}.

Fragments of first-order logic defined by the limitation of certain resources have been studied in detail. For instance, the quantifier alternation hierarchy, with its close relation with the dot-depth hierarchy of star-free languages, offers one of the oldest open problems in formal language theory: we know that the hierarchy is infinite and that its levels are characterized algebraically (by a property of the syntactic monoids), but we do not know whether these levels (besides levels 0 and 1) are decidable. In contrast, it is known that the quantifier alternation hierarchy for the first-order logic of the successor, $\FO[S]$, collapses at level 2 \cite{Thomas1982jcss,Pin2005dm}.

Another natural limitation considers the number of variables in a formula. This limitation has attracted a good deal of attention, as the
trade-off between formula size and number of variables is known to be
related with the trade-off between parallel time and number of
processes, see \cite{Immerman1999book,AdlerImmerman2003tcl,GroheSchweikardt2005lmcs}.

It is well-known that every first-order formula is equivalent to one using at most three variables. On the other hand, the first-order formulas using at most two variables, written $\FO^2[<]$, are strictly less expressive. The class of languages defined by such formulas admits many remarkable characterizations \cite{TessonTherien2002}. To begin with, a language is $\FO^2[<]$-definable if and only if it is recognized by a monoid in the pseudovariety $\DA$ \cite{TherienWilke1998stoc} (a precise definition will be given in Section~\ref{sec: varieties}). As with the characterization of $\FO[<]$-definability by aperiodic monoids, this characterization implies decidability.  The $\FO^2[<]$-definable languages are also characterized in terms of unambiguous products of languages (see Section~\ref{sec: unambiguous}) and in terms of the unary fragment of propositional temporal logic \cite{EtessamiVW2002ic} (see Section~\ref{sec: rankers and TL}).  For a survey of these properties, the reader is referred to \cite{TessonTherien2002,DiekertGK2008ijfcs}.

In this paper, we consider the quantifier alternation hierarchy within the two-variable fragment of first-order logic. We denote by $\FO^2_m[<]$ the fragment of $\FO^2[<]$ consisting of formulas using at most 2 variables and at most $m$ alternating blocks of quantifiers. In the sequel, we omit specifying the predicate $<$ and we write
simply $\FO$, $\FO^2$ or $\FO^2_m$.

Schwentick, Th\'erien and Vollmer introduced the so-called \emph{turtle programs} to characterize the expressive power of $\FO^2$ \cite{SchwentickTV2001dlt}. These programs are sequences of directional instructions of the form \textit{go to the next $a$ to the right, go to the next $b$ to the left}. More details can be found in Section~\ref{sec: rankers} below. Turtle programs were then used, under the name of \emph{rankers}, by Weis and Immerman \cite{WeisImmerman2009lmcs} (first published in \cite{WeisImmerman2007csl}) to characterize $\FO^2_m$ in terms of rankers with $m$ alternations of directions (right \textit{vs.} left). Their subtle characterization, Theorem~\ref{thm: IW refined} below, does not yield a decidability result. It forms however the basis of our results.

Rankers are actually better suited to the study of a natural alternation hierarchy within the unary fragment of propositional temporal logic (Sections~\ref{sec: rankers and TL} and~\ref{sec: Rm and unary}), than to the study of the quantifier alternation within $\FO^2[<]$. For the latter, we define the notion of \emph{condensed} rankers, which introduce a notion of efficiency in the path they describe in a word, see Section~\ref{sec: condensed rankers}.

Recent results of Kuf\-leitner and Lauser \cite{2012:KufleitnerLauser-fragments} and Straubing \cite{2011:Straubing} show that $\LangFO^2_m$ (the set of $\FO^2_m$-definable languages) forms a variety of languages. We show that the classes of languages defined by condensed rankers with at most $m$ changes of directions also form varieties of languages, written $\calR_m$ and $\calL_m$ depending on whether the initial move is towards the right or towards the left (Section~\ref{sec: Rm Lm varieties}). The meaning of these results is that membership of a language $L$ in these classes depends only on the syntactic monoid of $L$. This justifies using algebraic methods to approach the decidability problem for $\LangFO^2_m$ --- a technique that has proved very useful in a number of situations (see for instance \cite{Pin1986book,TessonTherien2002,StraubingWeil2011Hdbk,DiekertGK2008ijfcs}).

In fact, we use this algebraic approach to show that the classes $\calR_m$ and $\calL_m$ are decidable (Section~\ref{sec: structure Rm Lm}), and that they admit a neat characterization in terms of closure under alternated deterministic and co-deterministic products (Section~\ref{sec: deterministic}). Moreover, we show (Theorem~\ref{thm: interwoven}) that
$$\calR_m \cup \calL_m \subseteq \LangFO^2_m \subseteq \calR_{m+1} \cap \calL_{m+1}.$$
This shows that one can effectively compute, given a language $L$ in $\LangFO^2$, an integer $m$ such that $L$ is in $\LangFO^2_{m+1}$, possibly in $\LangFO^2_m$, but not in $\LangFO^2_{m-1}$. That is, we can compute the quantifier alternation depth of $L$ within one unit. As indicated above, this is much more precise than the current level of knowledge on the general quantifier alternation hierarchy in $\FO[<]$.

We conjecture that $\LangFO^2_m$ is actually equal to the intersection of $\calR_{m+1}$ and $\calL_{m+1}$. This would prove that each $\LangFO^2_m$ is decidable.

Many of these results were announced in \cite{KufleitnerWeilFO2mfcs}, with a few differences. In particular, the definition of the sets $\underR^\XX_{m,n}$ (Section~\ref{sec: rankers}) in \cite{KufleitnerWeilFO2mfcs} introduced a mistake which is corrected here. The proof of  \cite[Theorem 2]{KufleitnerWeilFO2mfcs} contained a gap: we do not have a proof that the classes defined by the alternation hierarchy within unary temporal logic are varieties. And the proof of \cite[Proposition 2.9]{KufleitnerWeilFO2mfcs} also contained a gap: the correct statement is Theorem~\ref{prop: agree 2 condensed} below.

\section{Rankers and logical hierarchies}

Let $A$ be a finite alphabet. We denote by $A^*$ the set of all words over $A$ (that is, of sequences of elements of $A$), and by $A^+$ the set of non-empty words.
If $u$ is a length $n$ ($n > 0$) word over $A$, we say that an integer $1\le i\le n$ is an \textit{$a$-position} of $u$ if the $i$-th letter of $u$, written $u[i]$, is an $a$.  If $1 \le i \le j \le n$, we let $u[i;j]$ be the factor $u[i]\cdots u[j]$ of $u$.

$\FO$ denotes the set of first-order formulas using the unary
predicates $\mathbf{a}$ ($a\in A$) and the binary predicate $<$, and $\FO^2$ denotes the fragment of $\FO$ consisting of formulas which use at most two variable symbols.

If $u$ is a length $n$ ($n > 0$) word over $A$, we identify the word $u$
with the logical structure $(\{1,\ldots, n\}, (\mathbf{a})_{a\in A})$,
where $\mathbf{a}$ denotes the set of $a$-positions in $u$.  Formulas
from $\FO$ are naturally interpreted over this structure, and we
denote by $L(\phi)$ the language \textit{defined by} the formula
$\phi\in\FO$, that is, the set of all words which satisfy $\phi$.

\subsection{Quantifier-alternation within $\FO^2$}

We now concentrate on $\FO^2$-formulas and we define two important
parameters concerning such formulas.  To simplify matters, we consider
only formulas where negation is used only on atomic formulas so that,
in particular, no quantifier is negated.  This is naturally possible
up to logical equivalence.  Now, with each formula $\phi\in \FO^2$, we
associate in the natural way a parsing tree: each occurence of a
quantification, $\exists x$ or $\forall x$, yields a unary node, each
occurrence of $\lor$ or $\land$ yields a binary node, and the leaves
are labeled with atomic or negated atomic formulas.  The
\textit{quantifier depth} of $\phi$ is the maximum number of
quantifiers along a path in its parsing tree.

With each path from root to leaf in this parsing tree, we also
associate its quantifier label, which is the sequence of quantifier
node labels ($\exists$ or $\forall$) encountered along this path.  A
\textit{block} in this quantifier label is a maximal factor consisting
only of $\exists$ or only of $\forall$, and we define the
\textit{number of blocks} of $\phi$ to be the maximum number of blocks
in the quantifier label of a path in its parsing tree. Naturally, the 
quantifier depth of $\phi$ is at least equal to its number of blocks.

We let $\FO^2_{m,n}$ ($n\ge m$) denote the set of first-order formulas
with quantifier depth at most $n$ and with at most $m$ blocks and let
$\FO^2_m$ denote the union of the $\FO^2_{m,n}$ for all $n$.  We also
denote by $\LangFO^2$ ($\LangFO^2_m$, $\LangFO^2_{m,n}$) the class of
$\FO^2$ ($\FO^2_m$, $\FO^2_{m,n}$)-definable languages.

\begin{rem}\label{FO21 and piecewise testable}
Recall that a language is \emph{piecewise testable} if it is a Boolean combination of languages of the form $A^*a_1A^*\cdots a_kA^*$ ($a_i\in A$). It is an elementary observation that the 
piecewise testable languages coincide with $\LangFO^2_1$.  It is well-known that this class of languages is decidable (see Section~\ref{sec: varieties} below).
\end{rem}

\subsection{Rankers}\label{sec: rankers}

A \textit{ranker} is a non-empty word on the alphabet $\{\XX_a,\YY_a \mid a\in A\}$. Rankers define positions in words: given a word $u \in A^+$ and a
letter $a\in A$, we denote by $\XX_a(u)$ (resp.\  $\YY_a(u)$) the least
(resp.\  greatest) integer $1\le i \le |u|$ such that $u[i] = a$.  If
$a$ does not occur in $u$, we say that $\YY_a(u)$ and $\XX_a(u)$ are
not defined.  If in addition $q$ is an integer such that $0 \le q \le
|u|$, we let
\begin{align*}
    \XX_a(u,q) &= q+\XX_a(u[q+1;|u|]) \\
    \YY_a(u,q) &= \YY_a(u[1;q-1]).
\end{align*}
These definitions are extended to all rankers: if $r'$ is a ranker,
$\ZZ \in \{\XX_a,\YY_a \mid a\in A\}$ and $r = r'\ZZ $, we let
$$r(u,q) = \ZZ(u,r'(u,q))$$
if $r'(u,q)$ and $\ZZ(u,r'(u,q))$ are defined, and we say that
$r(u,q)$ is undefined otherwise. In particular: rankers are processed from left to right.

Finally, if $r$ starts with an $\XX$- (resp.\  $\YY$-) letter, we say
that $r$ defines the position $r(u) = r(u,0)$ (resp.\  $r(u) =
r(u,|u|+1)$), or that it is undefined on $u$ if this position does not
exist. 

\begin{rem}
Rankers were first introduced, under the name of  \emph{turtle programs}, by Schwen\-tick, Th\'erien and Vollmer \cite{SchwentickTV2001dlt}, as  sequences of instructions: go to the next $a$ to the right, go to the next $b$ to the left, etc. These authors write $(\rightarrow,a)$ and $(\leftarrow,a)$ instead of $\XX_a$ and $\YY_a$. Weis and Immerman \cite{WeisImmerman2009lmcs} write $\RIGHT_a$ and $\LEFT_a$ instead, and they introduced the term \emph{ranker}.  We rather follow the notation in \cite{DiekertGK2008ijfcs,Kufleitner2007tcs,DiekertGastin2006ic}, where $\XX$ and $\YY$ refer to the future and past operators of \textsf{PTL}.
\end{rem}

\begin{exa}\label{first example ranker}
The ranker $\XX_a\YY_b\XX_c$ (go to the first $a$ starting from the left, thence to the first $b$ towards the left, thence to the first $c$ towards the right) is defined on $bac$ and $bca$, but not on $abc$ or $cba$.
\end{exa}

By $L(r)$ we denote the language of all words on which the ranker $r$ is
defined.  We say that the words $u$ and $v$ \textit{agree on a class}
$R$ of rankers if exactly the same rankers from $R$ are defined on $u$
and $v$.  And we say that two rankers $r$ and $s$ \textit{coincide} on
a word $u$ if they are both defined on $u$ and $r(u) = s(u)$.

\begin{exa}\label{rankers R1n}
If $r = \XX_{a_1}\cdots \XX_{a_k}$ (resp.\  $r = \YY_{a_k}\cdots \YY_{a_1}$), then $L(r)$ is the set of words that contain $a_1\cdots a_k$ as a subword, $L(r) = A^*a_1A^*\cdots a_kA^*$.
\end{exa}

The \textit{depth} of a ranker $r$ is defined to be its length as a word.
A \textit{block} in $r$ is a maximal factor in $\{\XX_a \mid
a\in A\}^+$ (an $\XX$-block) or in $\{\YY_a \mid a\in A\}^+$ (a
$\YY$-block).  If $n\ge m$, we denote by $R^\XX_{m,n}$ (resp.
$R^\YY_{m,n}$ ) the set of $m$-block, depth $n$ rankers, starting with
an $\XX$- (resp.\  $\YY$-) block, and we let $R_{m,n} = R^\XX_{m,n}
\cup R^\YY_{m,n}$ and $\underR^\XX_{m,n} = \bigcup_{m'\le m, n'\le
n}R^\XX_{m',n'} \cup \bigcup_{m'<m,n'<n}R^\YY_{m',n'}$.  We define
$\underR^\YY_{m,n}$ dually and we let $\underR^\XX_m = \bigcup_{n\ge
m}\underR^\XX_{m,n}$, $\underR^\YY_m = \bigcup_{n\ge
m}\underR^\YY_{m,n}$ and $\underR_m = \underR^\XX_m \cup
\underR^\YY_m$.

\begin{rem}
Readers familiar with \cite{WeisImmerman2009lmcs} will notice
differences between our $\underR^\XX_{m,n}$ and their analogous
$R^\star_{m\triangleright,n}$; introduced for technical reasons, it
creates no difference between our $\underR_{m,n}$ and their $R^\star_{m,n}$, the classes which intervene in crucial Theorem~\ref{thm: IW refined} below.
\end{rem}

\subsection{Rankers and unary temporal logic}\label{sec: rankers and TL}
Let us depart for a moment from the consideration of $\FO^2$-formulas,
to observe that rankers are naturally suited to describe the different
levels of a natural class of temporal logic.  The symbols $\XX_a$ and
$\YY_a$ ($a\in A$) can be seen as modal (temporal) operators, with the
\textit{future} and \textit{past} semantics respectively.  We denote
the resulting temporal logic (known as \textit{unary temporal logic})
by $\TL$: its only atomic formula is $\TRUE$, the other formulas are
built using Boolean connectives and modal operators.  Let $u\in A^+$
and let $0 \le i \le |u|+1$.  We say that $\TRUE$ holds at every
position $i$, $(u,i) \models \TRUE$; Boolean connectives are
interpreted as usual; and $(u,i) \models \XX_a\phi$ (resp.
$\YY_a\phi$) if and only if $(u,\XX_a(u,i)) \models \phi$ (resp.\ $(u,\YY_a(u,i)) \models \phi$).
We also say that $u \models \XX_a\phi$ (resp.
$\YY_a\phi$) if $(u,0) \models \XX_a\phi$ (resp.\  $(u,1+|u|) \models
\YY_a\phi$).

$\TL$ is a fragment of \textit{propositional temporal logic} $\PTL$;
the latter is expressively equivalent to $\FO$ and $\TL$ is
expressively equivalent to $\FO^2$ \cite{Kufleitner2007tcs}.

As in the case of $\FO^2$-formulas, one may consider the parsing tree
of a $\TL$-formula and define inductively its depth and number of
alternations (between past and future operators).  If $n\ge m$, the
fragment $\TL^\XX_{m,n}$ (resp.\  $\TL^\YY_{m,n}$) consists of the
$\TL$-formulas with depth $n$ and with $m$ alternated blocks, in which
every branch (of the parsing tree) with exactly $m$ alternations
starts with future (resp.\  past) operators. Branches with less alternations
may start with past (resp.\ future) operators.  The fragments
$\TL_{m,n}$, $\underTL^\XX_{m,n}$, $\underTL^\YY_{m,n}$,
$\underTL^\XX_{m}$, $\underTL^\YY_{m}$ and $\underTL_m$ are defined
according to the same pattern as in the definition of $R_{m,n}$,
$\underR^\XX_{m,n}$, $\underR^\YY_{m,n}$, $\underR^\XX_{m}$,
$\underR^\YY_{m}$ and $\underR_m$.  We also denote by
$\LangTL^\XX_{m,n}$ ($\LangTL^\XX_m$, $\underLangTL_m$, etc.)
the class of $\TL^\XX_{m,n}$ ($\TL^\XX_m$, $\TL_m$, etc.) -definable
languages.

\begin{prop}\label{prop: temporal}
    Let $1 \le m \le n$.  Two words satisfy the same $\underTL^\XX_{m,n}$
    formulas if and only if they agree on rankers from $\underR^\XX_{m,n}$.
    A language is in $\underLangTL^\XX_{m,n}$ if and only if it is a 
    Boolean combination of languages of the form $L(r)$, $r\in 
    \underR^\XX_{m,n}$.
    
    Similar statements hold for 
     $\underTL^\YY_{m,n}$, $\underTL^\XX_{m}$,
    $\underTL^\YY_{m}$ and $\underTL_m$, relative to the corresponding
    classes of rankers.
\end{prop}

\proof
Since every ranker can be viewed as a $\TL$-formula, it is easily verified that if $u$ and $v$ satisfy the same 
$\underTL^\XX_{m,n}$-formulas,
then they agree on rankers from 
$\underR^\XX_{m,n}$. 
To prove the converse, it suffices to show that a 
$\underTL^\XX_{m,n}$-formula
is equivalent to a Boolean combination of formulas that are expressed by a single ranker. That is: we only need to show that modalities can be brought outside the formula. This follows from the following elementary logical equivalences:
\begin{align*}
  \XX_a(\phi\wedge\psi) &\equiv \XX_a\phi \wedge \XX_a\psi, \\
  \XX_a(\phi\vee\psi) &\equiv \XX_a\phi \vee \XX_a\psi, \\
  \XX_a(\neg\phi) &\equiv \XX_a\TRUE \wedge \neg\XX_a\phi.
\end{align*}
\qed

\begin{rem}\label{TL1 is J}
    Together with Example~\ref{rankers R1n}, this proposition confirms
    the elementary observation that a language is $\underTL_1$ (resp.
    $\underTL_1^\XX$, $\underTL_1^\YY$) definable if and only if it is
    piecewise testable (see Remark~\ref{FO21 and piecewise testable}).  It follows that
    $\underTL_1$-definability is decidable.
\end{rem}

\subsection{Rankers and $\FO^2$}\label{sec: rankers and FO2}
The connection established by Weis and Immerman
\cite[Theorem 4.5]{WeisImmerman2009lmcs} between rankers and formulas in
$\FO^2_{m,n}$, Theorem~\ref{thm: IW refined} below, is much deeper.
If $x,y$ are integers, we let $\ord(x,y)$, the \textit{order type} of
$x$ and $y$, be one of the symbols $<$, $>$ or $=$, depending on
whether $x<y$, $x>y$ or $x=y$.

\begin{thm}[Weis and Immerman \cite{WeisImmerman2009lmcs}]\label{thm: IW refined}
    Let $u,v \in A^*$ and let $1\le m \le n$. Then $u$ and $v$ satisfy the 
    same formulas in $\FO^2_{m,n}$ if and only if
    \begin{itemize}
	\item[\textbf{(WI}]\textbf{1)}\enspace $u$ and $v$ agree on
	rankers from $\underR_{m,n}$,
	
	\item[\textbf{(WI}]\textbf{2)}\enspace if the rankers $r \in
	\underR_{m,n}$ and $r' \in \underR_{m-1,n-1}$ are defined on
	$u$ and $v$, then $\ord(r(u),r'(u)) = \ord(r(v),r'(v))$.
	
	\item[\textbf{(WI}]\textbf{3)}\enspace if $r \in
	\underR_{m,n}$ and $r' \in \underR_{m,n-1}$ are defined on $u$
	and $v$ and end with different direction letters, then
	$\ord(r(u),r'(u)) = \ord(r(v),r'(v))$.
    \end{itemize}
\end{thm}  

\begin{cor}\label{TLm in FO2m}
    For each $n\ge m\ge 1$, $\underLangTL_{m,n} \subseteq
    \LangFO^2_{m,n}$ and $\underLangTL_m \subseteq \LangFO^2_m$.
\end{cor}

\proof
Let $L$ be a $\underTL_{m,n}$-definable language.  For each $u\in L$,
let $\phi_u$ be the conjunction of the $\FO^2_{m,n}$-sentences
satisfied by $u$ and let $\phi$ be the disjunction of the formulas
$\phi_u$ ($u\in L$).  Since $\FO^2_{m,n}$ is finite (up to logical
equivalence), the conjunctions and disjunctions in the definition of
$\phi$ are all finite. We show that $L = L(\phi)$.

A word $v$ satisfies $\phi$ if and only if it satisfies $\phi_u$ for
some word $u\in L$.  Then $v$ satisfies the same
$\FO^2_{m,n}$-sentences as $u$ and, by comparing the statements in
Proposition~\ref{prop: temporal} and Theorem~\ref{thm: IW refined}, we
see that $u$ and $v$ satisfy the same $\underTL_{m,n}$-formulas.
Since $L$ is defined by such a formula, it follows that $v\in L$.
Conversely, every word $v\in L$ satisfies $\phi$ since it satisfies
$\phi_v$, which is logically equivalent to a term in the disjunction
defining $\phi$.  This concludes the proof.
\qed

\section{On varieties and pseudovarieties}\label{sec: varieties}

Recent results show that the $\FO^2_m$-definability of a language $L$ can be characterized algebraically, that is, in terms depending only on the syntactic monoid of $L$. This justifies exploring the algebraic path to tackle the decidability of this definability problem.
Eilenberg's theory of varieties provides the mathematical framework. In this section, we summarize the information on monoid and variety
theory that will be relevant for our purpose.  For more detailed
information and proofs, we refer the reader to
\cite{Pin1986book,Almeida1994book,TessonTherien2002,TessonTherien2007lmcs,StraubingWeil2011Hdbk},
among other sources.

A \textit{semigroup} is a set equipped with a binary associative
operation.  A \textit{monoid} is a semigroup which contains a
unit element.  The set $A^*$ of all words on alphabet $A$, equipped
with the concatenation product, is the \textit{free monoid} on $A$: it
has the specific property that, if $\phi\colon A\rightarrow M$ is a
map into a monoid, then there exists a unique monoid morphism
$\psi\colon A^* \rightarrow M$ which extends $\phi$.  Apart from free
monoids, the semigroups and monoids which we will consider in this
paper are finite.

If $A$ is a finite alphabet and $M$ is a finite monoid, we say that a
language $L\subseteq A^*$ is \textit{recognized} by $M$ if there
exists a morphism $\phi\colon A^* \rightarrow M$ such that $L =
\phi\inv(\phi(L))$.

\begin{exa}\label{example: [B]}
If $u\in A^*$ and $B\subseteq A$, let
\begin{align*}
    \Alpha(u) &= \{a\in A \mid u = vaw \textrm{ for some $v,w\in 
    A^*$}\},\\
    [B] &= \{u\in A^* \mid \Alpha(u) = B\}
\end{align*}
Let $\phi$ be the following morphism from $A^*$ into the direct
product of $|A|$ copies of the 2-element monoid 
$\{1,0\}$
(multiplicative): for each letter $a\in A$, $\phi(a)$ is the $A$-tuple in which every component is 
$1$, except for the $a$-component.
It is elementary to show that $[B] = \phi\inv(\phi([B]))$ and hence, $[B]$ is accepted by a monoid that is \textit{idempotent}
(every element is equal to its own square) and commutative.
Conversely, one can show that every language recognized by an idempotent and commutative
monoid is a Boolean combination of languages of the form $[B]$ ($B
\subseteq A$).
\end{exa}

A \textit{pseudovariety} of monoids is a class of
finite monoids which is closed under taking direct
products, homomorphic images and submonoids.  A
class of languages $\LangV$ is a collection $\LangV = (\LangV(A))_A$,
indexed by all finite alphabets $A$, such that $\LangV(A)$ is a set of
languages in $A^*$.  If $\V$ is a pseudovariety of monoids, we let
$\LangV(A)$ be the set of all languages of $A^*$ which are recognized
by a monoid in $\V$.  The class $\LangV$ has important closure
properties: each $\LangV(A)$ is closed under Boolean operations and
under taking residuals (if $L\in \LangV(A)$ and $u\in A^*$, then
$Lu\inv$ and $u\inv L$ are in $\LangV(A)$); and if $\phi\colon A^*
\rightarrow B^*$ is a morphism and $L \in \LangV(B)$, then
$\phi\inv(L) \in \LangV(A)$.  Classes of recognizable languages with
these properties are called \textit{varieties} of languages, and
Eilenberg's theorem (see \cite{Pin1986book}) states that the
correspondence $\V \mapsto \LangV$, from pseudovarieties of monoids to
varieties of languages, is one-to-one and onto.  Moreover, the
decidability of membership in the pseudovariety $\V$, implies the
decidability of the variety $\LangV$: indeed, a language is in
$\LangV$ if and only if its (effectively computable) syntactic monoid
is in $\V$.

For every finite semigroup $S$, there exists an integer, usually
denoted $\omega$, such that every element of the form $s^\omega$ in
$S$ is idempotent.  The \textit{Green relations} are another important
concept to describe semigroups and monoids: if $S$ is a semigroup and
$s,t\in S$, we say that $s\le_{\gJ}t$ (resp.\  $s \le_{\gR} t$, $s
\le_{\gL} t$) if $s = utv$ (resp.\  $s = tv$, $s = ut$) for some $u,v
\in S\cup\{1\}$.  We also say that $s \gJ t$ is $s\le_{\gJ} t$ and
$t\le_{\gJ} s$.  The relations $\gR$ and $\gL$ are defined similarly. 

Pseudovarieties that will be important in this paper are the 
following.

- $\Jone$, the pseudovariety of idempotent and commutative monoids; 
as discussed in Example~\ref{example: [B]}, the corresponding variety of languages consists 
of the Boolean combinations of languages of the form $[B]$.

- $\RR$, $\LL$ and $\J$, the pseudovarieties of $\gR$-, $\gL$- and 
$\gJ$-trivial monoids; a monoid is, say, $\RR$-trivial if each of its 
$\gR$-classes is a singleton. The variety of languages corresponding to $\J$ was described by Simon (see \cite{Pin1986book}): it is exactly the class of piecewise testable languages, i.e., the class of $\FO^2_1$-definable languages, see Remarks~\ref{FO21 and piecewise testable} and~\ref{TL1 is J}.

- $\Ap$, the variety of \textit{aperiodic} monoids, i.e., monoids in
which $x^\omega = x^{\omega+1}$ holds for each $x$.  Celebrated theorems of
Sch\"utzenberger, McNaughton and Papert and Kamp show that the
corresponding variety of languages consists of the star-free
languages, the languages that are definable in $\FO$, and the
languages definable in propositional temporal logic, see 
for instance \cite{Pin1986book,TessonTherien2007lmcs,DiekertGastin2008siwt,%
StraubingWeil2011Indianbk,StraubingWeil2011Hdbk}.

- $\DA$ is the pseudovariety of all monoids in which $(xy)^\omega x (xy)^\omega = (xy)^\omega$ for all $x,y$.  This pseudovariety has many characterizations in combinatorial, algebraic and logical terms. Of particular interest to us is the fact that the corresponding variety of languages consists of the languages that are definable in $\FO^2$, and equivalently, of the languages that are defined in unary temporal logic, see \cite{TessonTherien2002,TessonTherien2007lmcs,DiekertGK2008ijfcs,%
Kufleitner2007tcs,TrotterWeil1997au} among others.

- Straubing showed that, for each $m \ge 1$, $\LangFO^2_m$ is a variety of languages, and he described the corresponding pseudovariety of monoids, which we write $\VarFO^2_m$, in terms of iterated block products \cite{2011:Straubing}. We will not need to discuss the definition of the block product here, retaining only that this characterization does not imply decidability, and that Straubing gave identities (using products and $\omega$-powers like the identities given above for $\Ap$ and $\DA$) which he conjectures define each $\VarFO^2_m$. Establishing this conjecture would prove the decidability of $\FO^2_m$-definability.
\begin{enumerate}[$-$]
\item Kuf\-leitner and Lauser also showed that, for each 
$n \ge m \ge 1$, $\LangFO^2_{m,n}$ and $\LangFO^2_m$ form varieties of languages,
using a general result on \emph{logical fragments} \cite[Cor. 3.4]{2012:KufleitnerLauser-fragments}. Their result also does not imply a decidability statement.

\item On a given monoid $M$, we define the congruences $\sim_\K$ and $\sim_\D$ as follows.
\begin{iteMize}{$\bullet$}
\item $u \sim_\K v$ if and only if, for each idempotent $e$ in $M$, we have
either $eu,ev <_\calJ e$ or $eu = ev$,
\item $u \sim_\D v$ if and only if, for each idempotent $e$ in $M$, we have
either have $ue,ve <_\calJ e$ or $ue = ve$.
\end{iteMize}
\end{enumerate}
If $\V$ is a pseudovariety of monoids,  we say that the monoid $M \in \K\malcev\V$ if $M/{\sim_\K} \in \V$, and $M \in \D\malcev\V$ if $M/{\sim_\D} \in \V$. The classes $\K\malcev\V$ and $\D\malcev\V$ are pseudovarieties as well, which are usually defined in terms of Mal'cev products with the pseudovarieties $\K$ and $\D$, see \cite[Thm 4.6.50]{2009:RhodesSteinberg} or \cite{KrohnRT1965arbib,HallWeil1999sf}.

The following equalities are well-known \cite{Pin1986book}:
$$\K\malcev\Jone = \K\malcev\J = \RR,\enspace \D\malcev\Jone =
\D\malcev\J = \LL.$$

\section{Condensed rankers}\label{sec: condensed rankers}

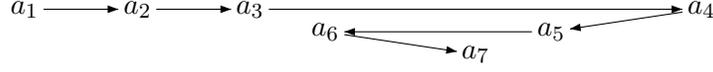
\begin{figure}
    \begin{picture}(60,20)(15,-10)
	\node[Nframe=n,Nw=5](n1)(0.0,-3.0){$a_1$}
	\node[Nframe=n,Nw=5](n2)(15.0,-3.0){$a_2$}
	\node[Nframe=n,Nw=5](n3)(30.0,-3.0){$a_3$}
	\node[Nframe=n,Nw=5](n4)(90.0,-3.0){$a_4$}
	\node[Nframe=n,Nw=5](n5)(70.0,-6.0){$a_5$}
	\node[Nframe=n,Nw=5](n6)(40.0,-6.0){$a_6$}
	\node[Nframe=n,Nw=5](n7)(60.0,-9.0){$a_7$}
	
	\drawedge(n1,n2){}
	\drawedge(n2,n3){}
	\drawedge(n3,n4){}
	\drawedge(n4,n5){}
	\drawedge(n5,n6){}
	\drawedge(n6,n7){}
    \end{picture}
    \caption{The positions defined by $r$ in $u$, when $r = \XX_{a_1}\XX_{a_2}\XX_{a_3}\XX_{a_4} \YY_{a_5}\YY_{a_6} \XX_{a_7}$ is condensed on $u$}\label{fig: condensed}
\end{figure}

Our main tool to approach the decidability of $\FO^2_m$-definability
lies in the notion of condensed rankers, a variant of rankers which
was introduced implicitly by Weis and Immerman to prove Theorem~\ref{fixed A}
below (see \cite[Theorem 4.7]{WeisImmerman2009lmcs}).  Recall that a
ranker can be seen as a sequence of directional instructions (see Example~\ref{first example ranker}).  We say that a
ranker $r$ is \emph{condensed on $u$} if it is defined on $u$, and if
the sequence of positions visited \textit{zooms in} on $r(u)$, never
crossing over a position already visited, see Figure~\ref{fig: condensed}.
Formally, $r = \ZZ_{1} \cdots \ZZ_{n}$ is condensed on $u$ if there
exists a chain of open intervals
$$(0;\abs{u}+1) = (i_0; j_0) \supset (i_1;j_1) \supset \cdots \supset
(i_{n-1};j_{n-1}) \ni r(u)$$
such that for all $1 \leq \ell \leq n-1$ the following properties are
satisfied:
\begin{iteMize}{$\bullet$}
    \item If $\ZZ_{\ell} \ZZ_{\ell + 1} = \XX_a \XX_b$ then
    $(i_\ell,j_\ell) = (\XX_a(u,i_{\ell-1}), j_{\ell-1})$.
    \item If $\ZZ_{\ell} \ZZ_{\ell + 1} = \YY_a \YY_b$ then
    $(i_\ell,j_\ell) = (i_{\ell-1}, \YY_a(u,j_{\ell-1})$.
    \item If $\ZZ_{\ell} \ZZ_{\ell+1} = \XX_a \YY_b$ then
    $(i_\ell,j_\ell) = (i_{\ell-1}, \XX_a(u,i_{\ell-1}))$.
    \item If $\ZZ_{\ell} \ZZ_{\ell+1} = \YY_a \XX_b$ then
    $(i_\ell,j_\ell) = (\YY_a(u,j_{\ell-1}), j_{\ell-1})$.
\end{iteMize}

\begin{rem}\label{iell are positions}
    The $i_\ell$ and $j_\ell$ are either 0 or $1+|u|$, or positions of
    the form $r'(u)$ for some prefix of $r'$ of $r$.  More precisely,
    if $r_\ell$ is the depth $\ell$ prefix of $r$ ($\ell < n$), then
    $r_\ell(u) = i_\ell$ if $\ZZ_{\ell+1}$ is of the form $\XX_a$, and
    $r_\ell(u) = j_\ell$ if $\ZZ_{\ell+1}$ is of the form $\YY_a$.
\end{rem}    

\begin{rem}\label{rk: follow direction next letter}
    If $r = r_1r_2$ is condensed on $u$, then $r(u) > r_1(u)$ if $r_2$ starts with an $\XX$-letter, and $r(u) < r_1(u)$ if $r_2$ starts with a $\YY$-letter.
\end{rem}    

\begin{exa}\label{example condensed}
    The ranker $\XX_a\YY_b\XX_c$ is defined on the words $bac$ and 
    $bca$, but it is condensed only on $bca$.
    
    Rankers in $\underR_{1,n}$ and rankers of the form
    $\XX_a\YY_{b_1}\cdots\YY_{b_k}$ or $\YY_a\XX_{b_1}\cdots\XX_{b_k}$
    are condensed on all words on which they are defined.
\end{exa}

Condensed rankers form a natural notion, which is equally well-suited
to the task of describing $\FO^2_m$-definability (see
Theorem~\ref{thm: IW refined condensed} below).  With respect to
$\TL$, for which Proposition~\ref{prop: temporal} shows a perfect
match with the notion of rankers, they can be interpreted as adding a
strong notion of unambiguity, see Section~\ref{sec: unambiguous} below
and the work of Lodaya, Pandya and Shah \cite{LodayaPS2008ifip}.

Let us say that two words $u$ and $v$ \textit{agree on condensed
rankers from a set} $R$ of rankers, if the same rankers in $R$ are
condensed on $u$ and $v$.  We write $u \RIGHT_{m,n} v$ (resp.\  $u
\LEFT_{m,n} v$) if $u$ and $v$ agree on condensed rankers in
$\underR^\XX_{m,n}$ (resp.\  $\underR^\YY_{m,n}$).

If $r$ is a ranker, let $L_c(r)$ be the language of all words on which $r$ is condensed. We define $\calR_m$ (resp.\  $\calL_m$) to be the Boolean algebra generated by the languages of the form $L_c(r)$, $r\in \underR^\XX_{m,n}$ (resp.\ $\underR^\YY_{m,n}$), $n\ge m$.

\subsection{Technical properties of condensed rankers}

A factorization $u = u_{-} a u_{+}$
of a word $u \in A^*$ is called the \textit{$a$-left factorization} of
$u$ if $a \not\in \Alpha(u_{-})$.  Symmetrically, $u = u_{-} a u_{+}$
is the \textit{$a$-right factorization} of $u$ if $a \not\in
\Alpha(u_{+})$.  Thus, the $a$-left (resp.\  $a$-right) factorization
of $a$ identifies the first occurrence of $a$ when reading $u$ from
the left (resp.\  the right).

Lemmas~\ref{condensed and factorizations} and~\ref{basics on rankers}
admit an elementary verification.  

\begin{lem}\label{condensed and factorizations}
    Let $s$ be a ranker, $a\in A$ and $r = \XX_as$.  Let also $u\in
    A^+$ and let $u = u_-au_+$ be its $a$-left factorization.  Then
    $r$ is condensed on $u$ if and only if
    \begin{iteMize}{$\bullet$}
	\item $s$ is condensed on $u_+$ if $s$ starts with an 
	$\XX$-block;
	
	\item $s$ is condensed on $u_-$ if $s$ starts with a
	$\YY$-block.
    \end{iteMize}
    A dual statement holds if $r$ is of the form $r = \YY_as$, with 
    respect to the $a$-right factorization of $u$.
\end{lem}    

\begin{lem}\label{basics on rankers}
    Let $r$ be a ranker and $a \in A$.  Let also $u\in A^+$ and let $u
    = u_-au_+$ be its $a$-left factorization.
    
    If $r$ starts with an $\XX$-letter, then
    \begin{iteMize}{$\bullet$}
	\item $r$ is defined on $u_-$ if and only if $r$ is defined 
	on $u$, $r$ does not contain $\XX_a$ or $\YY_a$ and, for 
	every prefix $p$ of $r$ ending with an $\XX$-letter, $p\YY_a$ 
	is not defined on $u$.
	
	\item $r$ is condensed on $u_-$
	\begin{iteMize}{$-$}
	    \item if and only if $r$ is defined on $u_-$ and condensed
	    on $u$,
	    
	    \item if and only if $r$ is condensed on $u$, $r$ does not
	    contain $\XX_a$ or $\YY_a$ and, if $p$ is the maximal
	    prefix of $r$ consisting only of $\XX$-letters, then
	    $p\YY_a$ is not defined on $u$,
	    
	    \item if and only if $r$ is condensed on $u$, $r$ does not
	    contain $\XX_a$ or $\YY_a$ and, if $p = \XX_{b_1}\cdots
	    \XX_{b_k}$ ($k \ge 1$) is the initial $\XX$-block of $r$,
	    then  $\XX_a\YY_{b_k}\cdots \YY_{b_1}$ is  defined on $u$.
	\end{iteMize}
	
	\item $r$ is defined on $u_+$ if and only if $\XX_ar$ is 
	defined on $u$ and, for every prefix $p$ of $r$ ending with a 
	$\YY$-letter, $\XX_ap\YY_a$ is defined on $u$.
	
	\item $r$ is condensed on $u_+$ if and only if $\XX_ar$ is 
	condensed on $u$.
    \end{iteMize}
    If $r$ starts with a $\YY$-letter, then
    \begin{iteMize}{$\bullet$}
	\item $r$ is defined on $u_-$ if and only if $\XX_ar$ is defined 
	on $u$, $r$ does not contain $\XX_a$ or $\YY_a$ and, for 
	every prefix $p$ of $r$ ending with an $\XX$-letter, $\XX_ap\YY_a$ 
	is not defined on $u$.
	
	\item $r$ is condensed on $u_-$ if and only if $\XX_ar$ is 
	condensed on $u$.

	\item $r$ is defined on $u_+$ if and only if $r$ is 
	defined on $u$ and, for every prefix $p$ of $r$ ending with a 
	$\YY$-letter, $p\YY_a$ is defined on $u$.
	
	\item $r$ is condensed on $u_+$
	\begin{iteMize}{$-$}
	    \item if and only if $r$ is defined on $u_+$ and condensed
	    on $u$,
	    
	    \item if and only if $r$ is condensed on $u$ and, if $p =
	    \YY_{b_1}\cdots \YY_{b_k}$ ($k \ge 1$) is the initial $\YY$-block
	    of $r$, then $p\YY_a$ is  defined on $u$.
	\end{iteMize}	
    \end{iteMize}
\end{lem}

\noindent We also note the following, very useful characterization of the
relations $\RIGHT_{m,n}$ and $\LEFT_{m,n}$.

\begin{prop}\label{prop: ppties congruences}
    The families of relations $\RIGHT_{m,n}$ and $\LEFT_{m,n}$ ($n
    \ge m \ge 1$) are uniquely determined by the following properties.
    \begin{enumerate}[\em(1)]
	\item $u \RIGHT_{1,n} v$ if and only if $u \LEFT_{1,n}
	v$, if and only if $u$ and $v$ have the same subwords of
	length at most $n$.
	
	\item If $m \ge 2$, then $u \RIGHT_{m,n} v$ if and only
	if $\Alpha(u) = \Alpha(v)$, $u \LEFT_{m-1,n-1} v$ and for each
	letter $a\in \Alpha(u)$, the $a$-left factorizations $u =
	u_-au_+$ and $v = v_-av_+$ satisfy $u_- \LEFT_{m-1,n-1} v_-$
	and $u_+ \RIGHT_{m,n-1} v_+$ ($u_+ \RIGHT_{m-1,n-1} v_+$ if $n = m$).
	
	\item If $m \ge 2$, then $u \LEFT_{m,n} v$ if and only if
	$\Alpha(u) = \Alpha(v)$, $u \RIGHT_{m-1,n-1} v$ and for each
	letter $a\in \Alpha(u)$, the $a$-right factorizations $u =
	u_-au_+$ and $v = v_-av_+$ satisfy $u_+ \RIGHT_{m-1,n-1} v_+$
	and $u_- \LEFT_{m,n-1} v_-$ ($u_- \LEFT_{m-1,n-1} v_-$ if $n = m$).
    \end{enumerate}
\end{prop}

\proof
Statement (1) follows directly from Examples~\ref{rankers R1n}
and~\ref{example condensed}.  Let us now assume that $m\ge 2$.

Suppose that $\Alpha(u) = \Alpha(v)$, $u \LEFT_{m-1,n-1} v$ and for
each $a\in \Alpha(u)$, the $a$-left factorizations $u = u_-au_+$ and
$v = v_-av_+$ satisfy $u_- \LEFT_{m-1,n-1} v_-$ and
$u_+ \RIGHT_{m,n-1} v_+$ if $n > m$ ($u_+ \RIGHT_{m-1,n-1} v_+$ if $n = m$).  Let $r\in \underR^\XX_{m,n}$ be condensed
on $u$.  If $r$ starts with a $\YY$-letter, then $r\in
\underR^\YY_{m-1,n-1}$, and hence $r$ is condensed on $v$ since $u
\LEFT_{m-1,n-1} v$.  If instead $r$ starts with an 
$\XX$-letter, say $r = \XX_as$, we consider the $a$-left factorizations of $u$ and $v$.  If
$s$ starts with a $\YY$-letter, then $s \in \underR^\YY_{m-1,n-1}$,
$s$ is condensed on $u_-$ (Lemma~\ref{condensed and factorizations})
and hence $s$ is condensed on $v_-$ since $u_- \LEFT_{m-1,n-1}v_-$,
from which it follows again that $r$ is condensed on $v$.  Finally, if
$s$ starts with an $\XX$-letter, then $s$ is condensed on $u_+$ by
Lemma~\ref{condensed and factorizations}. Moreover, $s\in
\underR^\XX_{m,n-1}$ if $n > m$. If $n=m$, we have in fact $r\in
\underR^\XX_{m-1,n}$ (since $r$ starts with two $\XX$-letters) and hence $s\in \underR^\XX_{m-1,n-1}$.  Since $u_+ \RIGHT_{m,n-1}
v_+$ if $n>m$ and $u_+ \RIGHT_{m-1,n-1}
v_+$ if $n = m$, it follows that $s$ is condensed on $v_+$, and hence $r$ is
condensed on $v$.

Conversely, let us assume that $u \RIGHT_{m,n} v$, that is, $u$ and
$v$ agree on condensed rankers in $\underR^\XX_{m,n}$.  Considering
rankers in $R^\XX_{1,1} \subseteq \underR^\XX_{m,n}$ shows that
$\Alpha(u) = \Alpha(v)$.  Similarly, considering rankers in
$\underR^\YY_{m-1,n-1} \subseteq \underR^\XX_{m,n}$ shows that $u
\LEFT_{m-1,n-1}v$.  Finally, let $a\in\Alpha(u)$ and let $u = u_-au_+$
and $v = v_-av_+$ be $a$-left factorizations.

Let $s\in \underR^\YY_{m-1,n-1}$ be condensed on $u_-$.  Note that $s$
contains neither $\XX_a$ nor $\YY_a$, since $a\not\in\Alpha(u_-)$.  If
$s$ starts with a $\YY$-letter, then $r = \XX_as$ is condensed on $u$
(Lemma~\ref{condensed and factorizations}) and since $r\in
\underR^\XX_{m,n}$, $r$ is condensed on $v$ as well, which implies
that $s$ is condensed on $v_-$.  If instead $s$ starts with an
$\XX$-letter, then $s$ is condensed on $u$ and hence on $v$.
Moreover, if $p = \XX_{b_1}\cdots \XX_{b_k}$ is the maximal prefix of
$s$ consisting only of $\XX$-letters, then $\XX_a\YY_{b_k}\cdots
\YY_{b_1} \in \underR^\XX_{2,n}$ is condensed on $u$
(Lemma~\ref{basics on rankers}). Since $\underR^\XX_{2,n} \subseteq \underR^\XX_{m,n}$, it is condensed on $v$ as well and hence,  $s$ is
condensed on $v_-$.

Finally, assume that $s\in\underR^\XX_{m,n-1}$ ($\underR^\XX_{m-1,n-1}$ if $n = m$) is
condensed on $u_+$.  The reasoning is similar: if $s$ starts with an
$\XX$-letter, then $\XX_as \in \underR^\XX_{m,n}$ is condensed on $u$.
Therefore $\XX_as$ is condensed on $v$ and $s$ is condensed on $v_+$.
If instead $s$ starts with a $\YY$-letter, then $s$ is condensed on $u$ and
$s\in\underR^\YY_{m-1,n-2}$ ($\underR^\XX_{m-2,n-2}$ if $n = m$). In particular $s\in \underR^\XX_{m,n}$ and hence, $s$ is condensed on
$v$ as well.  Moreover, if $p$ is the initial $\YY$-block of $s$, then
$p\YY_a$ is condensed on $u$.  Note that $p\YY_a \in
\underR^\YY_{1,n-1} \subseteq \underR^\XX_{m,n}$, so $p\YY_a$ is
condensed on $v$ and $s$ is condensed on $v_+$.
\qed

\begin{lem}\label{lem:simplepropd}\label{cor:leftANDrightNEW}
    Let $n \ge m \ge 2$, $u,v \in A^*$, $a\in A$  and let $u = u_{-} a u_{+}$ and $v = v_{-} a v_{+}$ be $a$-left
    factorizations.  If $u \RIGHT_{m,n}
    v$, then $u_{-} \RIGHT_{m,n-1} v_{-}$ ($u_{-} \RIGHT_{m-1,n-1} v_{-}$ if $n = m$). And if $u \LEFT_{m,n} v$, then $u_{+}
    \LEFT_{m,n-1} v_{+}$ ($u_{+}
    \LEFT_{m-1,n-1} v_{+}$ if $n = m$).  Dual
    statements hold for the factors of the $a$-right
    factorizations of $u$ and $v$ if $u \LEFT_{m,n} v$ or $u \RIGHT_{m,n} v$.
\end{lem}

\proof
We give the proof if $n > m$; it is easily adapted to the case where $n = m$.

Assume that $u \RIGHT_{m,n} v$ and $r\in \underR^\XX_{m,n-1}$ is condensed on $u_-$.  By Lemma~\ref{basics on rankers}, we have:

- If $r$ starts with an $\XX$-letter, then $r$ is condensed on $u$,
$r$ does not contain occurrences of $\XX_a$ or $\YY_a$, and if $p =
\XX_{b_1}\cdots\XX_{b_k}$ is the initial $\XX$-block of $r$, then $q =
\XX_a\YY_{b_k}\cdots\YY_{b_1}$ is condensed on $u$.  Since $q \in
R^\XX_{2,k+1}$ and $k < n$, we have $q \in \underR^\XX_{m,n}$ and
hence $r$ and $q$ are condensed on $v$.  Therefore $r$ is condensed on
$v_-$.

- If $r$ starts with a $\YY$-letter, then $\XX_ar$ is condensed on $u$.  But $r \in
\underR^\YY_{m-1,n-2}$, so $\XX_ar \in \underR^\XX_{m,n}$ and hence
$\XX_ar$ is condensed on $v$.  It follows that $r$ is condensed on
$v_-$.

Assume now that $u \LEFT_{m,n} v$ and $r \in \underR^\YY_{m,n-1}$ is condensed on $u_+$.  Then

- If $r$ starts with an $\XX$-letter (which is possible only if $m\ge
2$), then $r \in \underR^\XX_{m-1,n-2}$ and $\XX_ar$ is condensed on
$u$.  But $\XX_ar \in \underR^\XX_{m-1,n-1} \subseteq
\underR^\YY_{m,n}$, so $\XX_ar$ is condensed on $v$ and $r$ is
condensed on $v_+$.

- If instead $r$ starts with a $\YY$-letter, then $r$ is condensed on
$u$ and if $p$ is the initial $\YY$-block of $r$, then $p\YY_a$ is
condensed on $u$.  But $r,p\YY_a \in \underR^\YY_{m,n}$, so $r$ and
$p\YY_a$ are condensed on $v$, and $r$ is condensed on $v_+$.
\qed

\subsection{Condensed rankers, rankers and $\FO^2$}\label{sec: FO2m varieties}

We now show that, in the characterization of $\LangFO^2_{m,n}$ in
Theorem~\ref{thm: IW refined}, condensed rankers can be used just as
well.  This is done in Theorem~\ref{thm: IW refined condensed}.  The
first step is to relate agreement on rankers and agreement on
condensed rankers.  We start with a technical lemma.

\begin{lem}\label{technical defined condensed}
    If a ranker $r\in\underR^\ZZ_{m,n}$ ($\ZZ \in \{\XX,\YY\}$) is
    defined but not condensed on $u$, and if $s$ is the maximal prefix
    of $r$ which is condensed on $u$, then one of the following holds,
    for some $\ell\ge 1$:
    \begin{iteMize}{$\bullet$}
	\item $r = s\XX_bt$, $s = s_0 \YY_a \XX_{b_1}
	\cdots\XX_{b_{\ell-1}}$ and $s_0\YY_a(u) \le s(u) < s_0(u) \le
	s\XX_b(u)$;
	
	\item $r = s\YY_bt$, $s = s_0 \XX_a \YY_{b_1}
	\cdots\YY_{b_{\ell-1}}$ and $s_0\XX_a(u) \ge s(u) > s_0(u) >
	s\YY_b(u) = s_0\YY_b(u)$.
    \end{iteMize}
    Moreover $s_0$ is not empty, $s_0\in\underR^\ZZ_{m-1,n-\ell}$;
    $s\XX_b(u) = s_0(u)$ (resp.\  $s\YY_b(u) = s_0(u)$) if the last
    letter of $s_0$ is in $\{\XX_b, \YY_b\}$; and $s\XX_b(u) =
    s_0\XX_b(u)$  (resp.\  $s\YY_b(u) = s_0\YY_b(u)$) otherwise.
\end{lem}

\proof
Rankers in $R_{1,n}$ are condensed on each word on which they are defined (Example~\ref{example condensed}).  Therefore we have $m \ge 2$.

By hypothesis, $s\ne r$.  We consider the case where the first letter
after $s$ is an $\XX$-letter, the other case is dual.  Then $r$ is of
the form $r = s\XX_bt$, where $t$ may be empty.  In view of
Example~\ref{example condensed}, $s = s_0 \YY_{a}\XX_{b_1} \cdots
\XX_{b_{\ell-1}}$ for some non-empty $s_0$ and $\ell \ge 1$.  Since
$s$ is condensed on $u$ but $s\XX_b$ is not, we have the following (see Remark~\ref{rk: follow direction next letter}):
$$s_0\YY_a(u) < s_0\YY_a\XX_{b_1}(u) \cdots <
s_0\YY_a\XX_{b_1}\cdots\XX_{b_{\ell-1}}(u) = s(u) < s_0(u),$$
and $s\XX_b(u) \ge s_0(u)$.  More precisely, $s\XX_b(u)$ is the first
$b$-position to the right of $s(u)$, so $s\XX_b(u) = s_0(u)$ if $s_0(u)$
is a $b$-position (\textit{i.e.}, if $s_0$ ends with $\XX_b$ or
$\YY_b$), and $s\XX_b(u) = s_0\XX_b(u)$ otherwise.
\qed

\begin{prop}\label{fundamental condensed coincide}
    Let $n \ge m \ge 1$, $u,v \in A^+$ and $\ZZ \in \{\XX,\YY\}$.  If
    $u$ and $v$ agree on condensed rankers in $\underR_{m,n}^\ZZ$ and
    if $r\in \underR_{m,n}^\ZZ$ is defined on both $u$ and $v$, then
    there exists $r' \in \underR_{m,n}^\ZZ$ which is condensed on $u$
    and $v$ and coincides with $r$ on both words.
\end{prop}

\proof
The result is trivial if $m = 1$, since rankers in $R_{1,n}$ are
condensed on each word on which they are defined (Example~\ref{example
condensed}).  We now assume that $m \ge 2$.

Let $p$ and $q$ be positions in $u$ and $v$ and let $r \in
\underR_{m,n}^\ZZ$ such that $r(u) = p$ and $r(v) = q$.  If $r$ is not
condensed on $u$, then $r$ is not condensed on $v$ (since the two
words agree on condensed rankers).  With the notation of
Lemma~\ref{technical defined condensed}, $r$ coincides on both $u$ and
$v$ with $r' = s_0t$, $s_0\XX_bt$ or $s_0\YY_bt$ (depending on the
last letter of $s_0$ and on the letter following $s$ in $r$), which
starts with the same letter as $r$.  If $r'$ is not condensed on $u$
and $v$, we repeat the reasoning.  This process must terminate since
each iteration reduces the depth of $r'$.
\qed

\begin{prop}\label{fundamental condensed}
    Let $n \ge m \ge 1$, $u,v \in A^+$ and $\ZZ \in \{\XX,\YY\}$.  If
    $u$ and $v$ agree on condensed rankers in $\underR^\ZZ_{m,n}$,
    then they agree on rankers from the same class.
\end{prop}

\proof
If $u$ and $v$ do not agree on rankers from $\underR^\ZZ_{m,n}$, let
$r \in \underR_{m,n}^\ZZ$ be a minimum depth ranker on which $u$ and
$v$ disagree.  Without loss of generality, we may assume that $u \in
L(r)$ and $v\not\in L(r)$.  In particular, $r$ is not condensed on
$u$.

Let $s$, $s_0$ and $t$ be as in Lemma~\ref{technical defined
condensed}.  Without loss of generality again, we may assume that the
letter following $s$ in $r$ is $\XX_b$.  Since $s$ is condensed on $u$
and $s\XX_b$ is not, the ranker $s$ is condensed on $v$ and $s\XX_b$
is not.  Moreover, $s\XX_b$ coincides on $u$ with $s' = s_0$, or
$s_0\XX_b$, depending on the last letter of $s_0$.  Observe that $s'$
is shorter than $r$, so $s'$ is defined on $v$.  In particular, there
exists a $b$-position in $v$ to the right of $s$, which is not to the 
left of $s_0$ (since $s\XX_b$ is not condensed on $v$). It follows 
that $s\XX_b(v) = s'(v)$. Let now $r' = s't$: then $r'$ is shorter 
than $r$, it coincides with $r$ on $u$, and it is not defined on $v$ 
since $s'$ coincides with $s$ on that word. This contradicts the 
minimality of $r$.
\qed

We can now prove the following variant of Theorem~\ref{thm: IW
refined}.

\begin{thm}\label{thm: IW refined condensed}
    Let $u,v \in A^*$ and let $1\le m \le n$. Then $u$ and $v$ satisfy the 
    same formulas in $\FO^2_{m,n}$ if and only if
    \begin{enumerate}[\em\textbf{(WI 1c)}] 
	\item $u$ and $v$ agree on
	condensed rankers from $\underR_{m,n}$,
	
	\item[\em\textbf{(WI 2c)}] if the rankers $r \in
	\underR_{m,n}$ and $r' \in \underR_{m-1,n-1}$ are condensed on
	$u$ and $v$, then $\ord(r(u),r'(u)) = \ord(r(v),r'(v))$.
	
	\item[\em\textbf{(WI 3c)}] if $r \in
	\underR_{m,n}$ and $r' \in \underR_{m,n-1}$ are condensed on
	$u$ and $v$ and end with different direction letters, then
	$\ord(r(u),r'(u)) = \ord(r(v),r'(v))$.
    \end{enumerate}
\end{thm}  

\proof
We need to prove that together, Properties (\textbf{WI 1}),
(\textbf{WI 2}) and (\textbf{WI 3}) are equivalent to Properties
(\textbf{WI 1c}), (\textbf{WI 2c}) and (\textbf{WI 3c}).

Let us first assume that (\textbf{WI 1}), (\textbf{WI 2}) and
(\textbf{WI 3}) hold.  It is immediate that (\textbf{WI 2c}) and
(\textbf{WI 3c}) hold.  If (\textbf{WI 1c}) does not hold, let $r$ be
a ranker in $\underR_{m,n}$ which is condensed on $v$ and not on $u$.
Since (\textbf{WI 1}) holds, $r$ is defined on $u$.  Let $s_0$, $s$
and $t$ be as in Lemma~\ref{technical defined condensed} and let us
assume, without loss of generality, that the letter following $s$ in
$r$ is $\XX_b$.  Then $s_0$ and $s\XX_b$ are defined on both $u$ and
$v$, with $s_0\in\underR_{m-1,n-1}$ and $s\XX_b \in \underR_{m,n}$.
Since $r$ is condensed on $v$, we have $s\XX_b(v) < s_0(v)$, and since
$s\XX_b$ is not condensed on $u$, we have $s_0(v) \le s\XX_b(v)$,
contradicting Property (\textbf{WI 2}).  Thus (\textbf{WI 1c}) holds.

Conversely, let us assume that (\textbf{WI 1c}), (\textbf{WI 2c}) and
(\textbf{WI 3c}) hold.  Then (\textbf{WI 1}) holds by
Proposition~\ref{fundamental condensed}.  Let us verify Property
(\textbf{WI 2}): suppose that $r \in \underR_{m,n}$ and $r' \in
\underR_{m-1,n-1}$ are defined on $u$ and $v$.  In view of (\textbf{WI
1c}), Proposition~\ref{fundamental condensed coincide} shows that there
exist rankers $s\in \underR_{m,n}$ and $s'\in\underR_{m-1,n-1}$ which
are condensed on $u$ and $v$, and which coincide with $r$ and $r'$,
respectively, on both words.  By (\textbf{WI 2c}), we have
$\ord(s(u),s'(u)) = \ord(s(v),s'(v))$, and hence $\ord(r(u),r'(u)) =
\ord(r(v),r'(v))$.  Thus Property (\textbf{WI 2}) holds.  The
verification of (\textbf{WI 3}) is identical.
\qed
    
These results imply the following statement, which refines
Corollary~\ref{TLm in FO2m} and can be proved like that Corollary,
using Propositions~\ref{prop: temporal} and~\ref{fundamental
condensed}, and Theorem~\ref{thm: IW refined condensed}.

\begin{cor}\label{TLm in Rm in FO2m}
    For each $m \ge 1$, we have $\underLangTL^\XX_m \subseteq \calR_m
    \subseteq \LangFO^2_m$ and $\underLangTL^\YY_m \subseteq \calL_m
    \subseteq \LangFO^2_m$.
\end{cor}

\subsection{Condensed rankers determine a hierarchy of varieties}\label{sec: Rm Lm varieties}

We now examine the algebraic properties of the relations $\RIGHT_{m,n}$ and $\LEFT_{m,n}$.

\begin{lem}\label{right and left congruences}
    The relations $\RIGHT_{m,n}$ and $\LEFT_{m,n}$ are finite-index 
    congruences.
\end{lem}

\proof
The relations $\RIGHT_{m,n}$ and $\LEFT_{m,n}$ are clearly equivalence
relations, of finite index since $\underR_{m,n}$ is finite.  We now
verify that if $b\in A$ and if $u$ and $v$ are $\RIGHT_{m,n}$-equivalent, then so
are $ub$ and $vb$ (resp.\ $bu$ and $bv$).

The proof is by induction on $m+n$.  The property of having the same
subwords of length $n$ is easily seen to be a congruence (and the proof
of this fact can be found in \cite{Pin1986book} as it is related to
Simon's theorem on piecewise testable languages).  In view of
Proposition~\ref{prop: ppties congruences}~(1), this shows that
$\RIGHT_{1,n}$ and $\LEFT_{1,n}$ are congruences.

Let us now assume that $n\ge m\ge 2$ and $u \RIGHT_{m,n} v$.
By Proposition~\ref{prop: ppties congruences}~(2),
we have $\Alpha(u) = \Alpha(v)$ and $u
\LEFT_{m-1,n-1}v$.  It follows that
$\Alpha(ub) = \Alpha(bu) = \Alpha(vb) = \Alpha(bv)$, and that $ub \LEFT_{m-1,n-1} vb$ and $bu \LEFT_{m-1,n-1} bv$ by induction.

Let now $a\in \Alpha(u) \cup \{b\}$.  If $a\in \Alpha(u)$ and if $u =
u_-au_+$ and $v = v_-av_+$ are $a$-left factorizations, then the
$a$-left factorizations of $ub$ and $vb$ are $u_-\ a \ (u_+b)$ and
$v_-\ a \ (v_+b)$. And the $a$-left factorizations of $bu$ and $bv$ are $(bu_-)\ a \ u_+$ and
$(bv_-)\ a \ v_+$ --- unless $a = b$, in which case
these factorizations are $\epsilon b u$ and $\epsilon b v$.  By Proposition~\ref{prop: ppties congruences}~(2)
we have $u_- \LEFT_{m-1,n-1} v_-$ and $u_+ \RIGHT_{m,n-1}
v_+$ ($u_+ \RIGHT_{m-1,n-1}
v_+$ if $n = m$).  By induction, we have $u \RIGHT_{m,n-1}
v$, $bu_- \LEFT_{m-1,n-1} bv_-$ and $u_+b
\RIGHT_{m,n-1} v_+b$ ($u_+b \RIGHT_{m-1,n-1} v_+b$ if $n = m$).

If $a\not\in \Alpha(u)$, and hence $a = b$, the $a$-left factorizations of $ub$ and $vb$ (resp.\ $bu$ and $bv$) are $u\,b\,\epsilon$ and $v\,b\,\epsilon$ (resp.\ $\epsilon\,b\,u$ and $\epsilon\,b\,v$), and we do have $u \LEFT_{m-1,n-1} v$ and $u \RIGHT_{m,n-1} v$ ($u \RIGHT_{m-1,n-1} v$ if $m=n$).

Thus all the conditions in Proposition~\ref{prop: ppties congruences}~(2) are satisfied, whether $a$ occurs in $u$ and $v$ or not, and we have established that $ub \RIGHT_{m,n} vb$ and $bu \RIGHT_{m,n} bv$.  The proof regarding $\LEFT_{m,n}$ is symmetric.
\qed

\begin{lem}\label{fully invariant right and left congruences}
    If $\phi\colon A^* \to B^*$ is a morphism and if $u,v\in A^*$ are $\RIGHT_{m,n}$-equivalent (resp.\ $\LEFT_{m,n}$-equivalent), then so are $\phi(u)$ and $\phi(v)$.
\end{lem}

\proof
We carry out the proof for the congruence $\RIGHT_{m,n}$ by induction on $m+n$. The proof for $\LEFT_{m,n}$ is symmetrical.

For $m = 1$, we show that if a ranker $r \in \underR^\XX_{1,n}$ is condensed on $\phi(u)$, then it is condensed on $\phi(v)$. If $u = a_1\cdots a_\ell$, the word $\phi(u)$ has a natural factorization in blocks, namely the $\phi(a_i)$ and the sequence of positions in $\phi(u)$ defined by the prefixes of $r$ visits (some of) the $\phi(a_i)$-blocks. This yields a factorization of $r$, $r = r_1r_2 \cdots r_k$, where all the positions in $\phi(u)$ visited while running $r_1$ are in the same block, say, $\phi(a_{j(1)})$; then all the positions visited by the prefixes of $r$ between $r_1$ (excluded) and $r_1r_2$ (included) are in the block $\phi(a_{j(2)})$ with $j(2) > j(1)$; and so on. In particular, the ranker $\XX_{a_{j(1)}}\cdots \XX_{a_{j(k)}}$ is defined on $u$, and hence on $v$. Therefore $v = v_0 a_{j(1)} v_1 \cdots a_{j(k)} v_{k+1}$. By construction, each $r_i$ is defined on $\phi(a_{j(i)})$, so $r$ is defined on $\phi(v)$, and condensed on that word (Example~\ref{example condensed}).

We now let $n \ge m\ge 2$ and $u \RIGHT_{m,n} v$. It is immediate that $\Alpha(\phi(u)) = \Alpha(\phi(v))$ since $u$ and $v$ have the same alphabet. By Proposition~\ref{prop: ppties congruences}~(2), we have $u \LEFT_{m-1,n-1} v$, and by induction it follows that $\phi(u) \LEFT_{m-1,n-1} \phi(v)$. Let now $b\in \Alpha(\phi(u))$ and let $\phi(u) = x_-bx_+$ and $\phi(v) = y_-by_+$ be $b$-left factorizations.
The occurrence of $b$ thus singled out in $\phi(u)$ sits in some $\phi(a)$, $a\in A$, and the corresponding occurrence of $a$ in $u$ is the leftmost one: we have an $a$-left factorization $u = u_-au_+$ and a $b$-left factorization $\phi(a) = x'bx''$ such that $x_- = \phi(u_-)x'$ and $x_+ = x''\phi(u_+)$. Similarly, the leftmost occurrence of $b$ in $\phi(v)$ sits in some $\phi(a')$, $a'\in A$: $a'$ is the leftmost letter in $v$ such that $b$ occurs in $\phi(a')$. If $a' \ne a$, the consideration of the rankers $\XX_a\YY_{a'}$ and $\XX_{a'}\YY_a$, which are simultaneously defined or not defined on $u$ and $v$, yields a contradiction. Therefore $a' = a$ and if $v = v_-av_+$ is the $a$-left factorization, then $y_- = \phi(v_-)x'$ and $y_+ = x''\phi(v_+)$. By Proposition~\ref{prop: ppties congruences}~(2) again, we have $u_- \LEFT_{m-1,n-1} v_-$ and $u_+ \RIGHT_{m,n-1} v_+$ ($u_+ \RIGHT_{m-1,n-1} v_+$ if $n = m$). By induction, it follows that the same relations hold between the $\phi$-images of $u_-$, $v_-$, $u_+$ and $v_+$, and we have $x_- \LEFT_{m-1,n-1} y_-$ and $x_+ \RIGHT_{m,n-1} y_+$ ($x_+ \RIGHT_{m-1,n-1} y_+$ if $n = m$) by Lemma~\ref{right and left congruences}. Therefore $\phi(u) \RIGHT_{m,n} \phi(v)$ by Proposition~\ref{prop: ppties congruences}~(2).
\qed

For each $n \ge m \ge 1$, let $\RR_{m,n}$ (resp.\ $\LL_{m,n}$) be the pseudovariety of monoids generated respectively by the monoids of the form $A^*/\!\RIGHT_{m,n}$ (resp.\ $A^*/\!\LEFT_{m,n}$). Since $\RIGHT_{m,n'}$ refines $\RIGHT_{m,n}$ when $n' \ge n$, the sequence $(\RR_{m,n})_n$ is increasing and we let $\RR_m$ be its union (a pseudovariety as well).  The pseudovariety $\LL_m$ is defined similarly, as the union of the $\LL_{m,n}$.

\begin{cor}\label{Rm Lm free objects}
If $\gamma\colon A^* \to M$ is a morphism into a monoid in $\RR_{m,n}$, then there exists a morphism $\beta\colon A^*/\!\RIGHT_{m,n} \to M$ such that $\gamma = \beta\circ\pi_A$, where $\pi_A\colon A^*\to A^*/\!\RIGHT_{m,n}$ is the projection morphism. The same result holds for 
$\LL_{m,n}$ and the quotient $A^*/\!\LEFT_{m,n}$.
\end{cor}

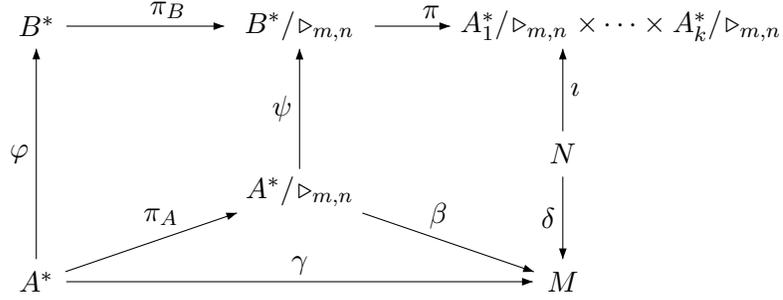
\begin{figure}
    \begin{picture}(85,37)(0,-37)
	\node[Nframe=n,NLangle=0.0,Nh=6.0,Nmr=2.5](n0)(0.0,-3.0){$B^*$}
	\node[Nframe=n,NLangle=0.0,Nh=6.0,Nw=20,Nmr=2.5](n1)(35.0,-3.0){$B^*/\!\RIGHT_{m,n}$}
	\node[Nframe=n,NLangle=0.0,Nh=6.0,Nw=30,Nmr=2.5](n2)(70.0,-3.0){\qquad\qquad$A_1^*/\!\RIGHT_{m,n} \times \cdots \times A_k^*/\!\RIGHT_{m,n}$}
	\node[Nframe=n,NLangle=0.0,Nh=6.0,Nw=20,Nmr=2.5](n3)(35.0,-25.0){$A^*/\!\RIGHT_{m,n}$}
	\node[Nframe=n,NLangle=0.0,Nh=6.0,Nmr=2.5](n4)(70.0,-20.0){$N$}
	\node[Nframe=n,NLangle=0.0,Nh=6.0,Nmr=2.5](n5)(0.0,-37.0){$A^*$}
	\node[Nframe=n,NLangle=0.0,Nh=6.0,Nmr=2.5](n6)(70.0,-37.0){$M$}
	
	\drawedge(n0,n1){$\pi_B$}
	\drawedge(n1,n2){$\pi$}
	\drawedge(n5,n0){$\phi$}
	\drawedge(n3,n1){$\psi$}
	\drawedge[ELside=r](n4,n2){$\imath$}
	\drawedge(n5,n3){$\pi_A$}
	\drawedge(n3,n6){$\beta$}
	\drawedge[ELside=r](n4,n6){$\delta$}
	\drawedge(n5,n6){$\gamma$}
    \end{picture}
    \caption{A commutative diagram}\label{fig: free objects}
\end{figure}

\proof
By definition, there exists an onto morphism $\delta\colon N \to M$, and an injective morphism $\imath\colon N \hookrightarrow A_1^*/\RIGHT_{m,n} \times \cdots \times A_k^*/\!\RIGHT_{m,n}$. Let $B$ be the disjoint union of the $A_i$, and for each $i$, let $\pi_i$ be the morphism from $B^*$ to $A_i^*$ which erases all the letters not in $A_i$. By Lemma~\ref{fully invariant right and left congruences}, $\RIGHT_{m,n}$-equivalent elements have $\RIGHT_{m,n}$-equivalent images, so we have a morphism $\pi\colon B^*/\!\RIGHT_{m,n} \to \prod_iA_i^*/\!\RIGHT_{m,n}$ as in Figure~\ref{fig: free objects}.

For each letter $a\in A$, we then pick a word $\phi(a)$ in $\pi_B\inv\pi\inv\delta\inv\gamma(a) \subseteq B^*$: this defines a morphism $\phi\colon A^*\to B^*$ such that $\delta\circ\pi\circ\pi_B\circ\phi = \gamma$. By Lemma~\ref{fully invariant right and left congruences} again, there exists a morphism $\psi\colon A^*/\!\RIGHT_{m,n} \to B^*/\!\RIGHT_{m,n}$ such that $\psi\circ\pi_A = \pi_B\circ\phi$. It follows that if $u\RIGHT_{m,n} v$, then $\pi_B\phi(u) = \pi_B\phi(v)$, and hence $\gamma(u) = \gamma(v)$. This concludes the proof.
\qed

\begin{cor}\label{cor: Rm Lm varieties}
    For each $m\ge 1$, $\calR_m$ and $\calL_m$ are varieties of
    languages and the corresponding pseudovarieties of monoids are
    $\RR_m$ and $\LL_m$.
\end{cor}

\proof
Every $L_c(r)$ ($r\in \underR^\XX_{m,n}$) is a union of
$\RIGHT_{m,n}$-classes, and hence it is recognized by
$A^*/\!\RIGHT_{m,n}$.  Therefore every language in $\calR_m$ is
recognized by a monoid in $\RR_m$ (and indeed, by $\pi_A\colon A^* \to A^*/\!\RIGHT_{m,n}$
for $n$ large enough).

Conversely, suppose that $L \subseteq A^*$ is recognized by a morphism $\gamma\colon A^* \to M$, into a monoid $M\in \RR_m$. Then $M \in \RR_{m,n}$ for some $n \ge m$ and by Corollary~\ref{Rm Lm free objects}, there exists a morphism $\beta\colon A^*/\!\RIGHT_{m,n} \to M$ such that $\gamma = \beta\circ\pi_A$. It follows that $L$ is also accepted by $\pi_A$, $L$ is a union of $\RIGHT_{m,n}$-classes, and hence $L \in \calR_m$.
\qed

\begin{exa}
    It follows from Proposition~\ref{prop: ppties congruences}~(i) 
    that $\calR_1 = \calL_1$ is the variety of piecewise testable 
    languages, and $\RR_1 = \LL_1 = \J$, the pseudovariety of 
    $\gJ$-trivial monoids.
\end{exa}

\begin{rem}
The proof of Corollary~\ref{cor: Rm Lm varieties} also establishes that for each $n \ge m\ge 1$, the Boolean algebra $\calR_{m,n}$ generated by the languages of the form $L_c(r)$, $r\in \underR^\XX_{m,n}$, defines a variety of languages, for which the corresponding pseudovariety of monoids is $\RR_{m,n}$. In variety-theoretic terms, Corollary~\ref{Rm Lm free objects} states that $A^*/\!\RIGHT_{m,n}$ is the free object of $\RR_{m,n}$ over the alphabet $A$. The symmetrical statement also holds for $\calL_{m,n}$ and the monoids $A^*/\!\LEFT_{m,n}$. 
\end{rem}

We note the following containments.

\begin{cor}\label{Rm, Lm in DA}
    For each $m \ge 1$, $\RR_m$ and $\LL_m$ are contained in $\DA$, and also in $\RR_{m+1} \cap \LL_{m+1}$.
\end{cor}

\proof
Since $\LangFO^2$ is the variety of languages corresponding to $\DA$, Corollary~\ref{TLm in Rm in FO2m} yields the containment of $\RR_m$ and $\LL_m$ in $\DA$. Similarly, $\calR_m$ and $\calL_m$ are contained in both $\calR_{m+1}$ and $\calL_{m+1}$ by definition of these classes of languages -- and this in turn implies the containment of the corresponding pseudovarieties.
\qed

\subsection{Condensed rankers and deterministic products}\label{sec: deterministic}

Recall that a product of languages $L = L_0a_1L_1\cdots a_kL_k$ ($k
\ge 1$, $a_i \in A$, $L_i \subseteq A^*$) is \textit{deterministic}
if, for $1 \le i \le k$, each word $u\in L$ has a unique prefix in
$L_0a_1L_1\cdots L_{i-1}a_i$.  If for each $i$, the letter $a_i$ does
not occur in $L_{i-1}$, the product $L_0a_1L_1\cdots a_kL_k$ is called
\textit{visibly deterministic}: this is obviously a particular case of
a deterministic product.

The definition of a \textit{co-deterministic} or \textit{visibly
co-deterministic} product is dual, in terms of suffixes instead of
prefixes.  If $\LangV$ is a class of languages and $A$ is a finite
alphabet, let $\LangV^{det}(A)$ (resp.\  $\LangV^{vdet}(A)$,
$\LangV^{codet}(A)$, $\LangV^{vcodet}(A)$) be the set of all Boolean
combinations of languages of $\LangV(A)$ and of deterministic (resp.
visibly deterministic, co-deterministic, visibly co-deterministic)
products of languages of $\LangV(A)$.

Pin gave algebraic characterizations of the 
operations $\LangV \longmapsto \LangV^{det}$ and $\LangV \longmapsto
\LangV^{codet}$, see \cite{1980:Pin,PinST1988jpaa}.

\begin{prop}\label{schutz}
    If $\LangV$ is a variety of languages and if $\V$ is the
    corresponding pseudovariety of monoids, then $\LangV^{det}$ and
    $\LangV^{codet}$ are varieties of languages
    and the corresponding pseudovarieties are, respectively, $\K\malcev\V$
    and $\D\malcev\V$. 
\end{prop}
    
This leads to the following statement.

\begin{thm}\label{malcev language statement}
    For each $m\ge 1$, we have $\calR_{m+1} = \calL_m^{vdet} =
    \calL_m^{det}$, $\calL_{m+1} = \calR_m^{vcodet} =
    \calR_m^{codet}$, $\RR_{m+1} = \K \malcev \LL_m$ and $\LL_{m+1} =
    \D \malcev \RR_m$.  In particular, $\RR_2 = \RR$ and $\LL_2 =
    \LL$.
\end{thm}

The proof uses the following technical property of monoids in $\DA$, 
whose proof can be found for instance in \cite[Lemma 4.2]{DiekertGK2008ijfcs}.

\begin{fact}\label{fact: DS}
    Let $\sigma\colon A^* \rightarrow S$ be a morphism into a monoid
    $S \in \DA$.  If $u,v\in A^*$, $a\in \Alpha(v)$ and $\sigma(u) \gR
    \sigma(uv)$, then $\sigma(uva) \gR \sigma(u)$.
\end{fact}

\proofof{Theorem~\ref{malcev language statement}}
It is immediate from the definition that $\calL_m^{vdet} \subseteq
\calL_m^{det}$.  

Let $u \in A^*$ and let $B = \Alpha(u)$.  For each $a\in B$, let $u =
u_-^{(a)}au_+^{(a)}$ be the $a$-left factorization of $u$.  Let $[B]$
be the language of all strings with alphabet $B$, $[B] = \{u \in A^*
\mid \Alpha(u) = B\}$.  Observe that
$$[B] = \bigcap_{a\in B} L_c(\XX_a) \setminus \bigcup_{a\not\in B} 
L_c(\XX_a) = \bigcap_{a\in B} L_c(\YY_a) \setminus \bigcup_{a\not\in B} 
L_c(\YY_a).$$
This shows that $[B]\in \calR_1 = \calL_1$.  (It is also well-known
that $[B]$ is piecewise testable, and hence $[B]\in \calR_1 =
\calL_1$.)

Now let $n > m \ge 1$.  It follows from Proposition~\ref{prop: ppties
congruences} that the $\RIGHT_{m+1,n}$-class of $u$ is the
intersection of $[B]$, the $\LEFT_{m,n-1}$-class of $u$ and the
products $KaL$ ($a \in B$) where $K$ is the $\LEFT_{m,n-1}$-class of
$u_-^{(a)}$ and $L$ is the
$\RIGHT_{m+1,n-1}$-class of $u_+^{(a)}$ if $n > m+1$, the
$\RIGHT_{m,n-1}$-class of $u_+^{(a)}$ if $n = m+1$.

By definition of an $a$-left factorization, each of these products is 
visibly deterministic and, since every $\LEFT_{m,n-1}$-class is a 
language in $\calL_m$, we have shown that the
$\RIGHT_{m+1,n}$-class of $u$ is in $\calL_m^{vdet}$. Thus 
$\calR_{m+1} \subseteq \calL_m^{vdet}$.

To establish the last inclusion, namely $\calL_m^{det}
\subseteq \calR_{m+1}$, we rather show $\K\malcev \LL_m \subseteq
\RR_{m+1}$.

Let $\gamma\colon A^* \rightarrow M$ be a surjective morphism, onto a
monoid $M \in \K\malcev\LL_m$: we want to show that there exists a
morphism from $A^*/\!\RIGHT_{m+1,n}$ onto $M$ for some $n > m$. 
Since
$M \in \K\malcev\LL_m$, the monoid $M/{\sim_\K} \in \LL_m$
and by Corollary~\ref{Rm Lm free objects}, there exists an integer $n$ and a morphism $\beta\colon A^*/{\LEFT_{m,n}} \to M/{\sim_\K}$ such that $\beta\circ\alpha = \pi\circ\gamma$, where $\alpha$ is the projection from $A^*$ onto $A^*/{\LEFT_{m,n}}$
and $\pi$ is the projection from $M$ onto $M/{\sim_\K}$, 
see Figure~\ref{fig: diagramme}.

\begin{figure}
    \begin{picture}(30,20)(0,-20)
	\node[Nframe=n,NLangle=0.0,Nh=6.0,Nmr=2.5](n0)(0.0,-3.0){$A^*$}
	\node[Nframe=n,NLangle=0.0,Nh=6.0,Nw=20,Nmr=2.5](n1)(30.0,-3.0){$A^*/{\LEFT_{m,n}}$}
	\node[Nframe=n,NLangle=0.0,Nh=6.0,Nmr=2.5](n3)(0.0,-20.0){$M$}
	\node[Nframe=n,NLangle=0.0,Nh=6.0,Nw=20,Nmr=2.5](n4)(30.0,-20.0){$M/{\sim_\K}$}
	
	\drawedge(n0,n1){$\alpha$}
	\drawedge(n1,n4){$\beta$}
	\drawedge[ELside=r](n0,n3){$\gamma$}
	\drawedge[ELside=r](n3,n4){$\pi$}
    \end{picture}
    \caption{$M \in \K\malcev \LL_m$}\label{fig: diagramme}
\end{figure}

Let $\ell$ be the maximal length of a strict $\gR$-chain in $M$, that
is: if $x_k <_{\gR} \ldots <_{\gR} x_1$ in $M$, then $k \le \ell$.  We
show that, for any $u,v\in A^*$,
\begin{align}\label{formula 1}
    u\RIGHT_{m+1,\ell|A| + n + 1} v &\Longrightarrow \gamma(u) =
    \gamma(v).
\end{align}

If $n' = \ell|A| + n + 1$, this implies the existence of a morphism
from $A^*/\!\RIGHT_{m+1,n'}$ onto $M$, as announced.

To prove implication~(\ref{formula 1}), it suffices to show that we
have
\begin{align}\label{formula 2}
    u\RIGHT_{m+1,\ell|\Alpha(u)| + n + 1} v &\Longrightarrow \gamma(u) = 
\gamma(v),
\end{align}
which we prove by induction on $|\Alpha(u)|$.  If $|\Alpha(u)| = 0$,
then $u = \epsilon$, $\Alpha(v) = \emptyset$ and $v = \epsilon$ as
well, so that $\gamma(u) = \gamma(v)$.

Now suppose that $u\ne \epsilon$ and assume that
$u\RIGHT_{m+1,\ell|\Alpha(u)| + n + 1} v$. Let $u = u_1a_1\cdots 
a_ku_{k+1}$ be the factorization of $u$ such that each $u_i$ is a 
word, each $a_i$ is a letter and
$$1 \gR \gamma(u_1) >_{\gR} \gamma(u_1a_1)
\cdots >_{\gR} \gamma(u_1a_1\cdots u_k a_k) \gR 
\gamma(u_1a_1\cdots a_ku_{k+1}).$$
Then $k+1 \le \ell$, so $k < \ell$.  Moreover, by Fact~\ref{fact: DS} (and Corollary~\ref{Rm, Lm in DA}),
for each $1\le i\le k$, $a_i \not\in u_i$, so that each product
$u_ia_i(u_{i+1}\cdots a_ku_{k+1})$ is an $a_i$-left factorization
($1\le i\le k$).

An easy induction on $k$, using Lemma~\ref{lem:simplepropd}, shows that $v$ can then be factored as
$$ v = v_1a_1v_2\cdots a_kv_{k+1},$$
where $u_i
\RIGHT_{m+1,\ell|\Alpha(u)|+n-i+1} v_i$ for each $1\le i \le k+1$.
Moreover, for $1\le i \le k$, $|\Alpha(u_i)| < |\Alpha(u)|$.  Since
$i\le k < \ell$, we have $\ell|\Alpha(u)|+n-i \ge
\ell|\Alpha(u_i)|+n+1$, and by induction, we have $\gamma(u_i) =
\gamma(v_i)$.  However, it is possible that $\Alpha(u_{k+1}) =
\Alpha(u)$, so we cannot conclude that $\gamma(u_{k+1}) = \gamma(v_{k+1})$.

But we do have the following:
$$u_{k+1} \RIGHT_{m+1,\ell|\Alpha(u)|+n-k} v_{k+1}\textrm{ and }
\gamma(u') = \gamma(v'),$$
where $u' = u_1a_1\cdots u_ka_k$ and $v' = v_1a_1\cdots v_ka_k$.
The first relation implies that $u_{k+1}$ and $v_{k+1}$ are $\LEFT_{m,\ell|\Alpha(u)|+n-k-1}$-equivalent.  Since $k<\ell$, we have $\ell|\Alpha(u)|+n-k-1 \ge n$, so $u_{k+1} \LEFT_{m,n} v_{k+1}$ and hence, $\pi\gamma(u_{k+1}) = \pi\gamma(v_{k+1})$, that is, $\gamma(u_{k+1}) \sim_\K \gamma(v_{k+1})$.

Moreover, there exists a string $x\in A^*$ such that
$\gamma(u') = \gamma(u'u_{k+1}x)$.  Let $\omega$ be an integer such
that every $\omega$-power is idempotent in $M$: then $\gamma(u') = \gamma(u')\gamma(u_{k+1}x)^\omega$.

Now observe that $\gamma(u_{k+1}x)^\omega \,\calJ\,\gamma(u_{k+1}x)^\omega\gamma(u_{k+1})$, since $\gamma(u_{k+1}x)^\omega = \gamma(u_{k+1}x)^{2\omega}$. It follows from $\gamma(u_{k+1}) \sim_\K \gamma(v_{k+1})$ that $\gamma(u_{k+1}x)^\omega\gamma(u_{k+1}) = \gamma(u_{k+1}x)^\omega\gamma(v_{k+1})$. Therefore we have
\begin{align*}
\gamma(u')\gamma(u_{k+1}) &= \gamma(u')\gamma(v_{k+1}) \quad\textrm{and hence} \\
\gamma(u) &= \gamma(u')\gamma(u_{k+1}) = \gamma(u')\gamma(v_{k+1}) = \gamma(v')\gamma(v_{k+1}) = \gamma(v).
\end{align*}
This concludes the proof of Formula~(\ref{formula 2}), and therefore
of Theorem~\ref{malcev language statement}.
\qedo

\subsection{Structure of the $\RR_m$ and $\LL_m$ hierarchies}\label{sec: structure Rm Lm}

It turns out that the hierarchies of pseudovarieties given by the
$\RR_m$ and the $\LL_m$ were studied in the semigroup-theoretic
literature (Trotter and Weil~\cite{TrotterWeil1997au}, Kuf\-leitner and Weil~\cite{KufleitnerWeilSF2010}).  In
\cite{KufleitnerWeilSF2010}, they are defined as the 
hierarchies of
pseudovarieties obtained from $\J$ by alternated applications of the
operations $\X\mapsto \K\malcev\X$ and $\X \mapsto \D\malcev\X$.
Theorem~\ref{malcev language statement} shows that these are the same
hierarchies as those considered in this paper%
\footnote{More precisely, the pseudovarieties $\RR_m$ and $\LL_m$ in
\cite{KufleitnerWeilSF2010} are pseudovarieties of semigroups,
and the $\RR_m$ and $\LL_m$ considered in this paper are the classes
of monoids in these pseudovarieties.}%
.  The following results are proved in
\cite[Section 4]{KufleitnerWeilSF2010}.

\begin{prop}\label{prop from sf}
    The hierarchies $(\RR_m)_m$ and $(\LL_m)_m$ are infinite chains of
    decidable pseudovarieties, and their unions are equal to $\DA$.
    Moreover, every $m$-generated monoid in $\DA$ lies in
    $\RR_{m+1}\cap\LL_{m+1}$.
\end{prop}

The results in \cite{TrotterWeil1997au,KufleitnerWeilSF2010} go actually further, and give defining pseudoidentities for the pseudovarieties $\RR_m$ and $\LL_m$.

\begin{rem}
The way up in the $\RR_m$-$\LL_m$ hierarchy, by means of Mal'cev products with $\K$ and $\D$, is strongly reminiscent of the structure of the lattice of band varieties \cite{Gerhard1970ja}. This observation is no coincidence, and forms the basis of the results in \cite{KufleitnerWeilSF2010} which are used here.
\end{rem}

\section{The $\RR_m$ hierarchy and unary temporal logic}\label{sec: Rm and unary}

We have seen in Corollary~\ref{TLm in Rm in FO2m} that $\underLangTL^\XX_m
\subseteq \calR_m$ and $\underLangTL^\YY_m \subseteq \calL_m$.  In
Theorem~\ref{prop: agree 2 condensed} below, we prove a weak 
converse. Let us however make the following observation.

\begin{prop}\label{TL2 = R2}
    We have
    \begin{align*}	
	\underLangTL^\XX_1 &= \underLangTL^\YY_1 = \calR_1 = 
	\calL_1,\\
	\underLangTL^\XX_2 &= \calR_2,\quad \underLangTL^\YY_2 =
	\calL_2.
    \end{align*}
\end{prop}

\proof
The statement concerning $\underLangTL_1$ was already proved in
Remark~\ref{TL1 is J}.  Let us now establish that $\calR_2 \subseteq
\underLangTL^\XX_2$.  We show, by induction on $n\ge 2$, that if $u$
and $v$ agree on rankers in $\underR^\XX_{2,2n}$, then they agree on
condensed rankers in $\underR^\XX_{2,n}$: $u \RIGHT_{2,n} v$.  We use
the characterization of $\RIGHT_{2,n}$ in Proposition~\ref{prop:
ppties congruences}.

The consideration of 1-letter rankers shows that $\Alpha(u) =
\Alpha(v)$.  Moreover, since $\underR^\YY_{1,n-1}$ is contained in
$\underR^\XX_{2,2n}$, and since these rankers are condensed where
they are defined, we find that $u \LEFT_{1,n-1} v$.  Similarly, let
$u = u_-au_+$ and $v = v_-av_+$ be $a$-left factorizations, 
and let $s \in \underR^\YY_{1,n-1}$.  Then $s$ is condensed on $u_-$ if and only if
$s$ is defined on $u_-$, if and only if $\XX_as$ is defined on $u$
(Lemma~\ref{basics on rankers}).  Since $\XX_as \in
\underR^\XX_{2,2n}$ and $u$ and $v$ agree on such rankers, it follows
that $\XX_as$ is defined on $v$, and $s$ is condensed on $v_-$.  Thus
$u_- \LEFT_{1,n-1} v_-$.

Now we need to show that $u_+ \RIGHT_{2,n-1} v_+$ if $n\ge 3$, $u_+ \RIGHT_{1,1} v_+$ if $n = 2$. Suppose first that $n = 2$ and consider $s\in \underR^\XX_{1,1}$, condensed on $u_+$. Then $s = \XX_b$ for some $b\in A$ and the consideration of $r = \XX_a\XX_b$ (in $\underR^\XX_{2,2}$) shows that $s$ is condensed on $v_+$ as well. This settles the case $n = 2$.

Let us now assume that $n \ge 3$ and let us show that $u_+
\RIGHT_{2,n-1} v_+$. By induction, it suffices to show that $u_+$
and $v_+$ agree on rankers in $\underR^\XX_{2,2n-2}$.  So let $s\in
\underR^\XX_{2,2n-2}$ be defined on $u_+$.  Then for every prefix $p$
of $s$ ending with a $\YY$-letter, $\XX_ap\YY_a$ is defined on $u$
(Lemma~\ref{basics on rankers}).  Since $\XX_ap\YY_a \in
\underR^\XX_{2,2n}$, it follows that $\XX_ap\YY_a$ is defined on $v$, 
and hence $s$ is defined on $v_+$. This concludes the proof.
\qed

Example~\ref{TL3 vs R3} below shows that the statement of Proposition~\ref{TL2 = R2} cannot be extended to the higher levels of the hierarchy.

\begin{exa}\label{TL3 vs R3}
    We show in this example that $\underLangTL_3^\XX$ is properly
    contained in $\calR_3$.  More precisely, let $r_0 =
    \XX_a\YY_b\XX_c \in R^\XX_{3,3}$.  We show that $L_c(r_0)$, a
    language in $\calR_3$, is not $\underTL^\XX_3$-definable.
    
    Let $u_n = (bc)^n(a(bc)^n)^n$ and $v_n = (bc)^nb(a(bc)^n)^n$ ($n
    \ge 1$).  It is easily verified that $r_0$ is condensed on $u_n$,
    and that it is defined and not condensed on $v_n$: that is, for
    each $n$, $u_n \in L_c(r_0)$ and $v_n \not\in L_c(r_0)$.
    
    We now show that $u_n$ and $v_n$ agree on all rankers in
    $\underR^\XX_{3,n}$, so that any $\underTL^\XX_3$-definable
    language contains either both $u_n$ and $v_n$, or neither -- and
    hence $L_c(r_0)$ is not $\underTL^\XX_3$-definable.
    
    Let $r\in \underR^\XX_{3,n}$.  If $r$ starts with a $\YY$-letter,
    then any two words ending with $(a(bc)^n)^n$ agree on $r$.  In
    particular, $u_n$ and $v_n$ agree on $r$.  Similarly, if $r$
    starts with an $\XX$-letter and does not contain the letters
    $\XX_a$ or $\YY_a$, then any two words starting with $(bc)^n$
    agree on $r$, so $u_n$ and $v_n$ agree on $r$.
	
    Finally, assume that $r$ starts with an $\XX$-letter and that $r =
    s_0\ZZ^{(1)}_a s_1\cdots \ZZ^{(k)}_a s_k$ with $k > 0$, each
    $\ZZ^{(i)} \in \{\XX,\YY\}$ and each $s_i$ a (possibly empty)
    ranker avoiding the letters $\XX_a$ and $\YY_a$.  We denote by
    $p_i$ the prefix $p_i = s_0 \ZZ^{(1)}_a s_1\cdots \ZZ^{(i)}_a$.
    
    Suppose first that $r\in R^\XX_{1,n}$.  Then $p_i$ coincides with
    $\XX_a^i$ on $u_n$ as well as on $v_n$.  Therefore $r$ is defined
    and coincides with $\XX_a^ks_k$ on both words.
    
    Suppose now that $r \in R^\XX_{2,n}$, say $r = r'r''$ with $r'$ a
    non-empty string of $\XX$-letters and $r''$ a non-empty string of
    $\YY$-letters.  If $r'$ is shorter than $p_1$, then $\ZZ^{(1)} =
    \YY$ and $r$ is not defined on either $u_n$ or $v_n$.  If
    $\ZZ^{(1)} = \XX$, let $i$ be maximal such that $p_i$ is a prefix
    of $r'$, say $r' = p_is'_i$.  Then $i > 0$ and $p_i$ coincides
    with $\XX_a^i$ on $u_n$, as well as on $v_n$.
    
    If $i = k$, then $r$ is defined on $u_n$ and $v_n$, and it
    coincides with $\XX_a^ks_k$ on both words.
    
    If $1\le i < k$, $s'_i$ is non-empty and $s_i$ is defined on
    $(bc)^n$, then $p_{i+1}$ coincides with $\XX_a^i$ on $u_n$ and
    $v_n$.  Thus $r$ is defined on $u_n$ (resp.\  $v_n$) if and only
    $i\ge k-i$, and in that case, it coincides with $\XX_a^{k-2i}s_k$.
    
    If $1 < i < k$, $s'_i$ is non-empty and $s_i$ is not defined on
    $(bc)^n$, or if $s'_i$ is empty, then $p_{i+1}$ coincides with
    $\XX_a^{i-1}$ on $u_n$ and $v_n$.  Thus $r$ is defined on $u_n$
    (resp.\  $v_n$) if and only $i > k-i$, and in that case, it
    coincides with $\XX_a^{k-2i-1}s_k$.
    
    Finally, if $1 = i < k$, $s'_i$ is non-empty and $s_i$ is not defined on
    $(bc)^n$, or if $s'_i$ is empty, then $r$ (and even $p_{i+1}$) is 
    not defined on either $u_n$ or $v_n$.
    
    Finally, let us assume that $r \in R^\XX_{3,n}$, say $r =
    r'r''r'''$ with $r'$ and $r'''$ non-empty strings of $\XX$-letters
    and $r''$ a non-empty string of $\YY$-letters.  Again, let $i$ be
    maximal such that $p_i$ is a prefix of $r'$ ($i=0$ if $p_1$ is not
    a prefix of $r'$) and let $j$ be maximal such that $p_j$ is a
    prefix of $r'r''$.  Then $r'r'' = p_js'_j$ for some prefix $s'_j$
    of $s_j$.  By the previous analysis, if $i \le 1 < j$, then
    $r'r''$ is not defined on $u_n$ nor on $v_n$, and hence neither is
    $r$.  In all other cases, $r'r''$ is defined on both words and
    coincides with $\XX_a^{2i-j}s'_j$ or $\XX_a^{2i-j-1}s'_j$.  Since
    $(k-j)+(2i-j)\le k$, $r$ is defined on $u_n$ and $v_n$, and
    coincides on these words with $\XX_a^{k-j+2i-j}s_k$ or
    $\XX_a^{k-j+2i-j-1}s_k$.
    
    To conclude this example, note that $u_n$ and $v_n$
    disagree on rankers in $\underR^\XX_4$.  More precisely, the
    ranker $\XX_a\YY_c\XX_b\YY_a$ is defined on $u_n$  but not on $v_n$.
Further getting ahead of ourselves, we note that this 
    example also shows (in view of Theorem~\ref{thm: interwoven}) 
    that $\underLangTL_3$ is properly contained in $\LangFO^2_3$.
\end{exa}

Finally we prove a result on the containment of the $\calR_m$ and $\calL_m$ hierarchies in the $ \underLangTL_m$ hierarchy. 

\begin{thm}\label{prop: agree 2 condensed}
    Let $m \ge 1$. Then $\calR_m \subseteq
    \underLangTL_{2m-1}^\XX$ and $\calL_m \subseteq
    \underLangTL_{2m-1}^\YY$.
    
    More precisely, for all $n \ge m$, $\ZZ\in \{\XX,\YY\}$ and
    $u,v \in A^*$, if $u$ and $v$ agree on rankers in
    $\underR^\ZZ_{2m-1,2n-1}$,
    then they agree on condensed rankers in $\underR^\ZZ_{m,n}$.
\end{thm}

\proof
Without loss of generality, we may assume $\ZZ = \XX$.  The proof is
by induction on $m$.  The result is trivial if $m = 1$, since $2m-1 =
1$ and $2n-1 \geq n$.  We now assume that $m \ge 2$ and $u$, $v$ agree
on rankers in $\underR^\XX_{2m-1,2n-1}$.

We use the characterization of $\RIGHT_{m,n}$ in
Proposition~\ref{prop: ppties congruences}: the consideration of
length 1 rankers shows that $\Alpha(u) = \Alpha(v)$.  Since
$\underR^\YY_{2m-3,2n-3}$ is contained in $\underR^\XX_{2m-1,2n-1}$,
we have $u \LEFT_{m-1,n-1} v$ by induction.  Now, for each letter
$a\in \Alpha(u)$, let $u = u_-au_+$ and $v = v_-av_+$ be the $a$-left
factorizations.  We want to show that $u_- \LEFT_{m-1,n-1} v_-$ and
$u_+ \RIGHT_{m,n-1} v_+$ ($u_+ \RIGHT_{m-1,n-1} v_+$ if $m = n$).  By induction, it suffices to show that
$u_-$ and $v_-$ agree on rankers in $\underR^\YY_{2m-3,2n-3}$, and
$u_+$ and $v_+$ agree on rankers in $\underR^\XX_{2m-1,2n-3}$
($\underR^\XX_{2m-3,2n-3}$ if $m = n$).  In the rest of the proof we
silently rely on the results of Lemma~\ref{basics on rankers}.

Let $s \in \underR^\YY_{2m-3,2n-3}$ be defined on $u_-$.  If $s$
starts with a $\YY$-block, then $\XX_as \in \underR^\XX_{2m-2,2n-2}$
and $\XX_as$ is defined on $u$.  Moreover, if $p$ is any prefix of
$s$, then $\XX_ap\YY_a \in \underR^\XX_{2m-1,2n-1}$ is not defined on
$u$.  It follows that $s$ is defined on $v_-$.

If instead $s$ starts with an $\XX$-block, then $s\in
\underR^\XX_{2m-4,2n-4}$ and $s$ is defined on $u$.  If $p$ is any
prefix of $s$, then $p\YY_a\in \underR^\XX_{2m-3,2n-3}$ and $p\YY_a$
is not defined on $u$.  As all these rankers are in
$\underR^\XX_{2m-1,2n-1}$, the same holds on $v$ and $s$ is defined
on $v_-$.

Let now $s \in \underR^\XX_{2m-1,2n-3}$ ($s \in
\underR^\XX_{2m-3,2n-3}$ if $n=m$) be defined on $u_+$.  If $s$ starts
with an $\XX$-block, then $\XX_a s \in \underR^\XX_{2m-1,2n-2}$
($\XX_a s \in \underR^\XX_{2m-3,2n-2}$ if $n=m$) and $\XX_a s$ is
defined on $u$.  Moreover, for each prefix $p$ of $s$ ending with a
$\YY$-letter, $\XX_ap\YY_a \in \underR^\XX_{2m-1,2n-1}$ ($\XX_ap\YY_a
\in \underR^\XX_{2m-3,2n-1}$ if $n=m$) and $\XX_ap\YY_a$
is defined on $u$.  As all these rankers are in
$\underR^\XX_{2m-1,2n-1}$, the same holds on $v$ and $s$ is defined on
$v_+$.

If instead $s$ starts with a $\YY$-block, then $s \in
\underR^\YY_{2m-2,2n-4}$ ($s \in \underR^\YY_{2m-4,2n-4}$ if $n=m$)
and $s$ is defined on $u$.  Moreover, if $p$ is any prefix of $s$
ending with a $\YY$-letter, $p\YY_a \in \underR^\YY_{2m-2,2n-3}$
($p\YY_a \in \underR^\YY_{2m-4,2n-3}$ if $n=m$) and $p\YY_a$ is
defined on $u$.  As all these rankers are in
$\underR^\XX_{2m-1,2n-1}$, the same holds on $v$ and $s$ is defined on
$v_+$.
\qed

The containment of $\calR_m$ and $\calL_m$ into
$\underLangTL^\XX_{2m-1}$ and
$\underLangTL^\YY_{2m-1}$, respectively, is not very precise,
unfortunately, especially in view of Theorem~\ref{fixed A}
below.

\section{The $\RR_m$ hierarchy and $\FO^2_m$}

The objective of this section is to prove the following theorem.

\begin{thm}\label{thm: interwoven}
    Let $m\ge 1$. Every language in $\calR_m$ or $\calL_m$ is 
    $\FO^2_m$-definable, and every $\FO^2_m$-definable language is in 
    $\calR_{m+1} \cap \calL_{m+1}$. Equivalently, we have
    $$\RR_m \vee \LL_m \subseteq \VarFO^2_m \subseteq \RR_{m+1} \cap 
    \LL_{m+1},$$
    where $\V\vee\W$ denotes the least pseudovariety containing $\V$ 
    and $\W$.
\end{thm}

\subsection{Are the containments in Theorem~\ref{thm: interwoven} strict?}

In the particular case where $m=1$, we know that $\RR_2 \cap \LL_2 =
\RR \cap \LL = \J = \RR_1\vee\LL_1$: this reflects the fact that $\LangFO^2_1$ is the class the piecewise testable languages.  However, we conjecture that this equality does not hold for larger values of $m$.

\begin{conj}
For $m\ge 2$, $\RR_m \vee \LL_m$ is properly contained in $\RR_{m+1} \cap \LL_{m+1}$.
\end{conj}

The following example proves the
conjecture for $m = 2$.

\begin{exa}\label{example R2L2 vs FO2}
    $L = \{b,c\}^*ca\{a,b\}^*$ is $\FO^2_2$-definable, by the
    following formula:
    \begin{eqnarray*}
	\exists i && (\mathbf{c}(i) \land (\forall j\ (j<i \to 
	\neg\mathbf{a}(j))) \land (\forall j\ (j>i \to 
	\neg\mathbf{c}(j)))) \\
	\land\enspace \exists i && (\mathbf{a}(i) \land (\forall j\ (j<i \to 
	\neg\mathbf{a}(j))) \land (\forall j\ (j>i \to 
	\neg\mathbf{c}(j))))\\
	\land\enspace \forall i && (\mathbf{b}(i) \to (\exists j\ (j<i \land 
	\mathbf{a}(j)) \lor (\exists j\ (j>i \land \mathbf{c}(j)))).
    \end{eqnarray*}
    The words $u_n = (bc)^n(ab)^n$ are in $L$, while the words $v_n =
    (bc)^nb(ca)^n$ are not.  Almeida and Azevedo showed that
    $\RR_2\vee\LL_2$ is defined by the pseudo-identity
    $(bc)^\omega(ab)^\omega = (bc)^\omega b(ab)^\omega$ \cite[Theorem
    9.2.13 and Exercise 9.2.15]{Almeida1994book}).  In particular, for
    each language $K$ recognized by a monoid in $\RR_2\vee\LL_2$, the
    words $u_n$ and $v_n$ (for $n$ large enough) are all in $K$, or
    all in the complement of $K$.  Therefore $L$ is not recognized by
    such a monoid, which proves that $\RR_2\vee\LL_2$ is strictly 
    contained in $\VarFO^2_2$, and hence also in $\RR_3\cap\LL_3$. It 
    also shows that $\underLangTL_2$ is properly contained in 
    $\LangFO^2_2$.
\end{exa}

Finally, we formulate the following conjecture.

\begin{conj}\label{conjecture FO2}
    For each $m\ge 1$, $\VarFO^2_m = \RR_{m+1} \cap \LL_{m+1}$.
\end{conj}

\subsection{Proof of Theorem~\ref{thm: interwoven}}

Corollary~\ref{TLm in Rm in FO2m} already established that every
language in $\calR_m$ or $\calL_m$ is $\FO^2_m$-definable\footnote{Of
course, the same fact can be proved by the direct construction of an
$\FO^2_m$-formula for each $\RIGHT_{m,n}$-class (by induction on $m$
and using Proposition~\ref{prop: ppties congruences}).}.

In view of Theorem~\ref{thm: IW refined condensed}, to establish that
$\LangFO^2_m$ is contained in $\calR_{m+1} \cap \calL_{m+1}$, it
suffices to prove the following result.

    \begin{quote}\it
	For each $n \ge m \ge 1$, if $u\RIGHT_{m+1,2n} v$
	or $u \LEFT_{m+1,2n} v$, then Properties (\textbf{WI 1c}),
	(\textbf{WI 2c}) and (\textbf{WI 3c}) hold for $m,n$.
    \end{quote}

The result is trivial if $m = 1$, since in that case, only Property
(\textbf{WI 1c}) is non-vacuous.

So we now assume that $m \ge 2$, and $u\RIGHT_{m+1,2n} v$ or $u
\LEFT_{m+1,2n} v$.  Property (\textbf{WI 1c}) holds trivially, by
definition of the $\RIGHT_{m+1,2n}$ and $\LEFT_{m+1,2n}$ relations.
We now concentrate on proving that Properties (\textbf{WI 2c}) and
(\textbf{WI 3c}) also hold for $m,n$, a task that will be completed 
in Section~\ref{sec: completing the proof}.

\subsubsection{The case where $r$ and $r'$ start with opposite directions}

\begin{prop}\label{lem:1ranker:1alphabet:differentdirections}
    Let $n \ge m \ge 1$, $r = \YY_{a_1 }s \in \underR_{m,n}^\YY$
    and $r' = \XX_c$.  If $u,v \in A^*$, $r$ is condensed on $u$ and
    $v$ and $u \RIGHT_{m,n+1} v$ or $u \LEFT_{m+1,n+1} v$, then
    $\ord(r(u),r'(u)) = \ord(r(v),r'(v))$.  The dual statement
    (involving $r = \XX_{a_1}s \in \underR_{m,n}^\XX$ and $r' =
    \YY_c$) holds as well.
\end{prop}    

\proof
\textit{First suppose that $u \LEFT_{m+1,n+1} v$,}\enspace
that is, $u$ and $v$ agree on condensed rankers in
$\underR^\YY_{m+1,n+1}$.  We are in exactly one of the following three
situations:

- $r\YY_c$ is defined on $u$, in which case $r'(u) < r(u)$;

- $r\YY_c$ is undefined on $u$ and $c$ is the last letter to occur in 
$r$, in which case  $r'(u) = r(u)$;

- $r\YY_c$ is undefined on $u$ and $c$ is not the last letter to occur in 
$r$, in which case $r(u) < r'(u)$. 

The same trichotomy holds for $v$.  Since $r\YY_c \in
\underR^\YY_{m+1,n+1}$,  $u$ and $v$ agree on $r\YY_c$
(Proposition~\ref{fundamental condensed}), and hence $\ord(r(u),r'(u))
= \ord(r(v),r'(v))$.

\medskip

\noindent\textit{Let us now assume that $u \RIGHT_{m,n+1} v$,}\enspace
so that $u$ and $v$ agree on condensed rankers in
$\underR^\XX_{m,n+1}$.  If $m = 1$ then $r$ is of the form $r =
\YY_{a_1} \cdots \YY_{a_k}$ and we observe again that

- either $\XX_c \XX_{a_k} \cdots \XX_{a_1} \in R_{1,n+1}^\XX$ is
defined on $u$, and we have $r'(u) < r(u)$;

- or  $\XX_c \XX_{a_k} \cdots \XX_{a_1}$ is undefined on
$u$ and $c = a_k$, and we have $r'(u) = r(u)$;

- or $\XX_c \XX_{a_k} \cdots \XX_{a_1}$ is undefined on
$u$ and $c \not= a_k$, and we have $r'(u) > r(u)$.

The same holds for $v$ since $\XX_c \XX_{a_k} \cdots \XX_{a_1} \in
\underR^\XX_{1,n+1}$ and such rankers are condensed where they are
defined.  Therefore we have $\ord(r(u),r'(u)) = \ord(r(v),r'(v))$.

We now assume that $m \geq 2$.  Let $u = u_- c u_+$ and $v = v_- c
v_+$ be $c$-left factorizations. We distinguish two cases depending 
on the direction of the second letter of $r$.

First suppose that $r = \YY_{a_1} \YY_{a_2} s'$.  If $a_1 \not\in
\Alpha(u_+)$, then $r(u) < r'(u)$ (because $r$ is condensed on $u$).
Since $u_+ \RIGHT_{m,n} v_+$, we have $\Alpha(u_+) = \Alpha(v_+)$,
so $r(v) < r'(v)$ as well.  If instead $a_1 \in \Alpha(u_+) =
\Alpha(v_+)$, let $u_+ = u_0 a_1 u_1$ and $v_+ = v_0 a_1 v_1$ be the
$a_1$-right factorizations.  Then
\begin{eqnarray*}
    \ord(r(u),r'(u)) &=& \ord(\YY_{a_2}s'(u_- c u_0), r'(u_- c u_0))\textrm{ and}\cr
    \ord(r(v),r'(v)) &=& \ord(\YY_{a_2}s'(v_- c v_0), r'(v_- c v_0)).
\end{eqnarray*}
Since $(u_- c u_0)a_1u_1$ and $(v_- c
v_0)a_1u_1$ are $a_1$-right factorizations as well, we deduce from
Lemma~\ref{cor:leftANDrightNEW} that $u_- c u_0 \RIGHT_{m,n}
v_- c v_0$ and it follows by induction on the length of $r$ that
\begin{eqnarray*}
    \ord(\YY_{a_2} s'(u_- c u_0), r'(u_- c u_0)) &=& \ord(\YY_{a_2}
    s'(v_- c v_0), r'(v_- c v_0)).
\end{eqnarray*}

The other case is $r = \YY_{a_1} \XX_{b_1} s'$.  If $a_1 \in \Alpha(c
u_+) = \Alpha(c v_+)$ then $r'(u) < r(u)$ and $r'(v) < r(v)$.  If
instead $a_1 \not\in \Alpha(c u_+) = \Alpha(c v_+)$, we first consider
the case where $r$ has a single alternation, i.e., $r = \YY_{a_1}
\XX_{b_1} \cdots \XX_{b_{k}}$.  We have $r(u) < r'(u)$ if and only if
$r$ is defined on $u_-$, and hence condensed (Example~\ref{example
condensed}).  Since $u_- \RIGHT_{m,n} v_-$
(Lemma~\ref{lem:simplepropd}), this is the case if and only if $r$ is
defined on $v_-$.  Hence, if $r$ is defined on $u_-$, we have $r(u) <
r'(u)$ and $r(v) < r'(v)$.  If $r$ is not defined on $u_-$, but
$\YY_{a_1} \XX_{b_1} \cdots \XX_{b_{k-1}}$ is defined on $u_-$ and
$b_k = c$, then the same holds for $v$ and we have $r(u) = r'(u)$ and
$r(v) = r'(v)$.  Otherwise, we have $r(u) > r'(u)$ and $r(v) > r'(v)$.

The last situation arises if $r$ is of the form $r = \YY_{a_1}
\XX_{b_1} \cdots \XX_{b_{k}} \YY_d s''$.  In particular, $m \ge 3$.
If $\YY_{a_1} \XX_{b_1} \cdots \XX_{b_{k}}$ is defined on $u_- c$,
then it is defined on $v_-c$ as well (by the same reasoning as in the
previous paragraph) and we have $r(u) < r'(u)$ and $r(v) < r'(v)$.

Similarly, if $\YY_{a_1} \XX_{b_1} \cdots \XX_{b_{k-1}}$ is not
defined on $u_-$ and $v_-$, then we have $r'(u) < r(u)$ and $r'(v) <
r(v)$.

Finally, let us assume that $\YY_{a_1}
\XX_{b_1} \cdots \XX_{b_{k}}$ is not defined on $u_- c$ or $v_- c$,
but $\YY_{a_1} \XX_{b_1} \cdots \XX_{b_{k-1}}$ is defined on $u_-$ and
$v_-$.  Let $u_+ = u_0 b_k u_1$ and $v_+ = v_0 b_k v_1$ be $b_k$-left
factorizations.  Then
\begin{eqnarray*}
    \ord(r(u),r'(u)) &=& \ord(\YY_d s''(u_- c u_0),r'(u_- c
    u_0))\textrm{ and}\\
    \ord(r(v),r'(v)) &=& \ord(\YY_d s''(v_- c v_0),r'(v_- c v_0)).
\end{eqnarray*}
Since $u \RIGHT_{m,n+1} v$, we have $u_+ \RIGHT_{m,n} v_+$, and by
Lemma~\ref{lem:simplepropd} , $u_- \RIGHT_{m,n} v_-$ and $u_0
\RIGHT_{m,n-1} v_0$.  Therefore $u_- c u_0 \RIGHT_{m,n-1} v_- c v_0$.
Since $\YY_d s'' \in \underR^\YY_{m-2,n-2}$ is condensed on both $u_-
c u_0$ and $v_- c v_0$, we conclude by induction on the length of $r$
that $\ord(\YY_d s''(u_- c u_0),r'(u_- c u_0)) = \ord(\YY_d s''(v_- c
v_0),r'(v_- c v_0))$ and hence $\ord(r(u),r'(u)) = \ord(r(v),r'(v))$.

This concludes the proof.
\qed

\begin{prop}\label{lem:2rankers:differentdirections}
    Let $n > m \ge 1$, let $r = \XX_as \in \underR^\XX_m$ and $r' =
    \YY_bs' \in \underR^\YY_m$ such that $|r|+|r'| \le n$, and let
    $u,v \in A^*$ such that $r$ and $r'$ are condensed on $u$ and $v$.
    If $u \RIGHT_{m+1,n} v$ or $u \LEFT_{m+1,n} v$, then
    $\ord(r(u),r'(u)) = \ord(r(v),r'(v))$.
\end{prop}   

\proof
Without loss of generality, we assume that $u \RIGHT_{m+1,n} v$.  We
proceed by induction, first on $m$.  If $m = 1$, then $r =
\XX_{a_1}\cdots \XX_{a_k}$ and $r' = \YY_{b_1}\cdots \YY_{b_\ell}$
with $k+l\le n$.  We observe that if $p =
r\XX_{b_\ell}\cdots\XX_{b_1}$ is defined on $u$, then $r(u) < r'(u)$;
if $p$ is not defined on $u$, but $a_k = b_\ell$ and
$r\XX_{b_{\ell-1}}\cdots\XX_{b_1}$ is defined on $u$, then $r(u) =
r'(u)$; and in all other cases, $r(u) > r'(u)$.  The same holds for
$v$, and this completes the proof in case $m = 1$.

We now assume that $m\ge 2$ and proceed by induction on $n$.  We first
note that if one of $r$, $r'$ has length $1$, then the result was
established in
Proposition~\ref{lem:1ranker:1alphabet:differentdirections}.  We now
assume that $\abs{r}, \abs{r'} \geq 2$ (so $\abs{r}, \abs{r'} \leq n-2$).

Suppose that $n = m+1$ and let $\beta(r)$ the number of alternating
blocks in $r$: then $\beta(r) \le |r| \le n - |r'| \le n-2 = m-1$.
The same inequality holds for $r'$ and we conclude by induction on
$m$.

We must now consider the case where $n > m+1 > 2$.  In particular, we
have $r\in \underR^\XX_{m,n-2}$ and $r'\in \underR^\XX_{m+1,n-1}$.

\medskip

\noindent\textit{First case: $s$ starts with an $\XX$-block.}\enspace
Let $u = u_- a u_+$ and $v = v_- a v_+$ be $a$-left-factorizations.
Then $s$ is condensed on $u_+$ and $v_+$ and $u_+ \RIGHT_{m+1,n-1}
v_+$, so $u_+$ and $v_+$ agree on rankers in $\underR^\XX_{m+1,n-1}$
(Proposition~\ref{fundamental condensed}).  In particular, $u_+$ and
$v_+$ agree on $r'$.  If $r'$ is defined on $u_+$, then
$\ord(r(u),r'(u)) = \ord(s(u_+),r'(u_+))$.  Moreover, $r'$ is defined
on $v_+$ as well and $\ord(r(v),r'(v)) = \ord(s(v_+),r'(v_+))$, so we
conclude by induction.  If instead $r'$ is not defined on $u_+$ or
$v_+$, then $r'(u) \le \XX_a(u) < r(u)$ and $r'(v) \le \XX_a(v) <
r(v)$.

\medskip

\noindent\textit{Second case: $s'$ starts with a $\YY$-block.}\enspace
Let $u = u_- b u_+$ and $v = v_- b v_+$ be $b$-right factorizations.
Then $u_- \RIGHT_{m+1,n-1} v_-$ by Lemma~\ref{cor:leftANDrightNEW} and
this case can be handled exactly like the previous one.

\medskip

\noindent\textit{Third case: $s$ starts with a $\YY$-block and $s'$
starts with an $\XX$-block.}\enspace
If $\XX_a(u) \leq \YY_b(u)$, then $\XX_a(v) \leq \YY_b(v)$ (by
Proposition~\ref{lem:1ranker:1alphabet:differentdirections}), we have
$r(u) < \XX_a(u) \leq \YY_b(u)< r'(u)$, and the same inequalities hold
for $v$.

We now assume that $\XX_a(u) > \YY_b(u)$ and $\XX_a(v) > \YY_b(v)$.
In particular, $a \neq b$.  Identifying the first $a$ and the last $b$
in $u$ and $v$, we get factorizations $u = u_- b u_0 a u_+$ and $v =
v_- b v_0 a v_+$ such that $a \not\in \Alpha(u_- b u_0) \cup
\Alpha(v_- b v_0)$ and $b \not\in \Alpha(u_0 a u_+) \cup \Alpha(v_0 a
v_+)$.  In particular, $r(u) = s(u_-bu_0)$, $r'(u)$ is the position
$s'(u_0au_+)$ in the suffix $u_0au_+$ of $u$, and the same holds in
$v$.  Moreover, $u = (u_-bu_0)au_+$ is an $a$-left factorization, $u =
u_-b(u_0au_+)$ is a $b$-right factorization, and the same holds in
$v$.  Therefore, and since $u \RIGHT_{m+1,n} v$, we have
$u_-bu_0 \LEFT_{m,n-1} v_-bv_0$ by definition and $u_0 \LEFT_{m,n-2}
v_0$ by Lemma~\ref{lem:simplepropd}.

Since $s\in \underR^\YY_{m-1,n-3}$ and $s'\in \underR^\XX_{m-1,n-3}
\subseteq \underR^\YY_{m,n-2}$, Proposition~\ref{fundamental
condensed} shows that, if $s$ is not defined on $u_0$, then it is not
defined on $v_0$ either, and $r(u) \le \YY_b(u) < r'(u)$ and
similarly, $r(v) < r'(v)$.  Symmetrically, if $s'$ is not defined on
$u_0$, then $r(u) < \XX_a(u) \le r'(u)$ and $r(v) < r'(v)$.

Finally, if $s$ and $s'$ are defined on $u_0$, then
\begin{eqnarray*}
    \ord(r(u),r'(u)) &=& \ord(s(u_0),s'(u_0))\textrm{ and}\\
    \ord(r(v),r'(v)) &=& \ord(s(v_0),s'(v_0)),
\end{eqnarray*}
and we conclude by induction.
\qed

\subsubsection{The case where $r$ and $r'$ start with the same direction}

\begin{prop}\label{lem:1ranker:1alphabet:samedirection}
    Let $n \ge m \ge 2$, $r \in \underR_{m,n}^\XX$ starting with an
    $\XX$-letter, and $r' = \XX_c$.  If $u,v \in A^*$, $r$ is
    condensed on $u$ and $v$ and $u \RIGHT_{m,n+1} v$, then
    $\ord(r(u),r'(u)) = \ord(r(v),r'(v))$.  The dual statement
    (involving $r\in \underR_{m,n}^\YY$, $r' = \YY_c$ starting with a
    $\YY$-letter, and $u \LEFT_{m,n+1} v$) holds as well.
\end{prop}

\proof
We proceed by induction, first on $m$.  If $m = 2$, then either $r =
\XX_{a_1}\cdots\XX_{a_k}$ or $r =
\XX_{a_1}\cdots\XX_{a_k}\YY_{b_1}\cdots\YY_{b_\ell}$.  In the first
case, the order type $\ord(r(u),r'(u))$ depends, as in the proof of
Proposition~\ref{lem:2rankers:differentdirections}, on whether
$\XX_c\YY_{a_k}\cdots \YY_{a_1}$ is defined on $u$, or if it is not
defined, whether $a_k = c$ and $\XX_c\YY_{a_{k-1}}\cdots \YY_{a_1}$ is
defined.  Since these rankers are in $\underR^\XX_{2,n+1}$ and are
condensed where they are defined (Example~\ref{example condensed}), we
have $\ord(r(u),r'(u)) = \ord(r(v),r'(v))$.

In the second case, where $r =
\XX_{a_1}\cdots\XX_{a_k}\YY_{b_1}\cdots\YY_{b_\ell}$, three cases
arise: if $r\YY_c$ is defined on $u$, then $\XX_c(u) < r(u)$; if
$r\YY_c$ is not defined and $c = b_\ell$, then $\XX_c(u) = r(u)$; in
all other cases, $r(u) < \XX_c(u)$.  Since $u \RIGHT_{2,n+1} v$ and
$r\YY_c\in \underR^\XX_{2,n+1}$, Proposition~\ref{fundamental
condensed} shows that $r\YY_c$ is defined on $u$ if and only if it is
defined on $v$, and $\ord(r(u),r'(u)) = \ord(r(v),r'(v))$.

We now assume that $m \ge 3$. If $r$ has less than $m$ alternating 
blocks, we conclude by induction on $m$. Let us suppose now that $r$ 
has $m$ alternating blocks and let us proceed by induction on $|r| \ge m$.

Let $r = \XX_as$. If $s$ starts with a $\YY$-letter (which includes the base case where
$|r| = m$), then $s\in \underR^\YY_{m-1,n-1}$ is condensed on $u_-$ and
$v_-$.  If $c \not\in \Alpha(u_-) = \Alpha(v_-)$, then $r(u) < r'(u)$
and $r(v) < r'(v)$.  In all other cases,
\begin{eqnarray*}
\ord(r(u),r'(u)) &=& \ord(s(u_-),r'(u_-))\textrm{ and}\\
\ord(r(v),r'(v)) &=& \ord(s(v_-),r'(v_-)).
\end{eqnarray*}
Since $u_- \RIGHT_{m,n} v_-$ by Lemma~\ref{lem:simplepropd}, these two
order types are equal by
Proposition~\ref{lem:1ranker:1alphabet:differentdirections}.

If instead $s$ starts with an $\XX$-letter, then $|r| > m$, $s\in
\underR^\XX_{m,n-1}$ is condensed on $u_+$ and $v_+$
(Lemma~\ref{condensed and factorizations}) and we distinguish two
cases.  If $c \in \Alpha(u_-a) = \Alpha(v_-a)$, then $r'(u) < r(u)$
and $r'(v) < r(v)$.  Otherwise
\begin{eqnarray*}
    \ord(r(u),r'(u)) &=& \ord(s(u_+),r'(u_+))\textrm{ and}\\
    \ord(r(v),r'(v)) &=&    \ord(s(v_+),r'(v_+)).
\end{eqnarray*}
Since $u_+ \RIGHT_{m,n} v_+$, these two order types are equal by
induction on $n$.
\qed

\begin{prop}\label{lem:tworankers:samedirection}
  Let $n \ge m \ge 2$, let $r = \XX_as \in \underR^\XX_m$ and $r' =
  \XX_bs' \in \underR^\XX_{m-1}$ such that $|r|+|r'| \le n$, and let
  $u,v \in A^*$ such that $r$ and $r'$ are condensed on $u$ and $v$.
  If $u \RIGHT_{m,n} v$, then $\ord(r(u),r'(u)) = \ord(r(v),r'(v))$.
  The dual statement (where $r,r'$ start with $\YY$-blocks and $u
  \LEFT_{m,n} v$) holds as well.
\end{prop}

\proof
The proof is by induction on $m$, and then on $n$.  If one of $r$ and
$r'$ has length $1$, then the result was established in
Proposition~\ref{lem:1ranker:1alphabet:samedirection}.  This takes
care of the cases where $n \le 3$, including the base case $m = n =
2$.  We now assume that $\abs{r}, \abs{r'} \geq 2$.

Let us observe that under this assumption, if $n = m$, then the number
of alternating blocks in $r$ is less than or equal to $m-2$: indeed it
is at most equal to $|r| \le n-2 = m-2$.  The same inequality holds
for $r'$, so this situation is handled by induction on $m$.  We can
now assume that $n > m$.

Let $u = u_- a u_+ = u'_- b u'_+$ and $v = v_- a v_+ = v'_- b v'_+$ be
$a$-left and $b$-left factorizations.

\medskip

\noindent\textit{First case: $a = b$.}\enspace
If $s$ starts with an $\XX$-block and $s'$ starts with a $\YY$-block,
then $r'(u) < r(u)$ and $r'(v) < r(v)$.  Dually, if $s$ starts with a
$\YY$-block and $s'$ starts with an $\XX$-block, then $r'(u) > r(u)$
and $r'(v) > r(v)$.

If $s$ and $s'$ both start with a $\YY$-block (which can happen only
if $m-1 \ge 2$), then $s\in \underR^\YY_{m-1}$ and $s'\in
\underR^\YY_{m-2}$ are condensed on $u_-$ and $v_-$ and
\begin{eqnarray*}
\ord(r(u),r'(u)) &=& \ord(s(u_-),s'(u_-))\textrm{ and}\\
\ord(r(v),r'(v)) &=& \ord(s(v_-),s'(v_-)).
\end{eqnarray*}
Since $u_- \LEFT_{m-1,n-1} v_-$ and $|s|+|s'| \le n-2$, we have
$\ord(r(u),r'(u)) = \ord(r(v),r'(v))$ by induction on $m$.

If instead $s$ and $s'$ both start with an $\XX$-block, then $s\in
\underR^\XX_m$ and $s' \in \underR^\XX_{m-1}$ are condensed on $u_+$
and $v_+$, and we have
\begin{eqnarray*}
\ord(r(u),r'(u)) &=& \ord(s(u_+),s'(u_+))\textrm{ and}\\
\ord(r(v),r'(v)) &=& \ord(s(v_+),s'(v_+)).
\end{eqnarray*}
Since $u_+ \RIGHT_{m,n-1} v_+$  and $|s|+|s'| \le n-2$, we have 
$\ord(r(u),r'(u)) = \ord(r(v),r'(v))$ by induction on $n$.

\medskip

\noindent\textit{Second case: $a \ne b$, $s$ and $s'$ start with
$\XX$-blocks.}\enspace
Then $s\in \underR^\XX_m$ and $s'\in \underR^\XX_{m-1}$ are condensed
on $u_+$ and $v_+$.  Without loss of generality, $\XX_b(u) <
\XX_a(u)$, so we have $r(u) = \XX_as(u) = \XX_b\XX_as(u) = \XX_br(u)$.
In particular, $\ord(r(u),r'(u)) = \ord(r(u'_+),s'(u'_+))$.  By
Proposition~\ref{lem:1ranker:1alphabet:samedirection}, we also have
$\XX_b(v) < \XX_a(v)$, and hence $\ord(r(v),r'(v)) =
\ord(r(v'_+),s'(v'_+))$.  Since $u \RIGHT_{m,n} v$, we have $u'_+
\RIGHT_{m,n-1} v'_+$ and we conclude by induction on $n$ since $|r| +
|s'| \le n-1$.

\medskip

\noindent\textit{Third case: $a \ne b$, $s$ and $s'$ start with
$\YY$-blocks.}\enspace
This can occur only if $m-1 \ge 2$.  Then $s \in \underR^\YY_{m-1}$
and $s' \in \underR^\YY_{m-2}$ are condensed on $u_-$ and $v_-$, $r(u)
= s(u_-)$ and $r'(u) = s'(u'_-)$, and the same equalities hold for
$v$.  Without loss of generality, we may assume that $\XX_b(u) <
\XX_a(u)$, and hence $\XX_b(v) < \XX_a(v)$
(Proposition~\ref{lem:1ranker:1alphabet:samedirection}).  Let $u_0$
and $v_0$ be such that $u = u'_-bu_0au_+$ and $v = v'_-bv_0av_+$: then
$u_0$ is the left factor in the $a$-left decomposition of $u'_+$ and
the right factor in the $b$-left decomposition of $u_-$.  An analogous
statement is true for $v_0$.  There are two cases, depending on
whether $s$ is defined on $bu_0$.  If this is the case, then $r'(u) <
r(u)$.  Moreover, we have $u'_+ \RIGHT_{m,n-1} v'_+$ and $u_0
\LEFT_{m-1,n-2} v_0$, so $s$ is defined on $bv_0$ as well, by
Proposition~\ref{fundamental condensed}.

If instead, $s$ is not defined on $bu_0$ or $bv_0$, let $p$ be the
longest prefix of $s$ which is defined on $bu_0$ (and hence on
$bv_0$): then $p$ is either empty or a $\YY$-block and $s = p \YY_{c}
t$, where $c$ has no occurrence in $u[\XX_b(u);\XX_a p(u)-1]$ (so
$\YY_c$ is defined on $u'_-$).

If $\YY_{c} t$ is defined on $u'_-$, then $r(u) = s(u_-) = \YY_c
t(u'_-)$, so that
$$\ord(r(u),r'(u)) = \ord(\YY_ct(u_-),s'(u_-)).$$
Now $u \RIGHT_{m,n} v$ implies $u'_- \RIGHT_{m,n-1} v'_-$ by
Proposition~\ref{lem:simplepropd}, so $\YY_ct$ is defined on $v'_-$
and hence we have $\ord(r(v),r'(v)) = \ord(\YY_c t(v'_-),s'(v'_-))$ as
well.  Since $|\YY_c t| \le |s| < |r|$, we conclude by induction that
$\ord(r(u),r'(u)) = \ord(r(v),r'(v))$.

If $\YY_{c} t$ is not defined on $u'_-$, then let $\YY_cq$ be the
longest prefix of $\YY_{c} t$ which is defined on $u'_-$ (and hence on
$v'_-$).  Then $q$ is either empty or an $\XX$-block and $\YY_{c} t =
\YY_{c}q \XX_{d} t'$.  If $d = b$, then $q\XX_d(u'_-b) = \XX_b(u)$, so
$r(u) = \XX_b t' (u)$ and similarly, $r(v) = \XX_b t' (v)$.  We
conclude by induction on $m$ that $\ord(r(u),r'(u)) =
\ord(r(v),r'(v))$ since $\XX_b t'$ has 2 blocks less than $r$.

If $d \ne b$, then we have $\XX_b(u) < \XX_a p \YY_c q \XX_d(u)$. If $\XX_dt'$ 
is defined on $bu_0$, then $r(u)$ lies in $u_0$ and $r'(u)$ lies in 
$u'_-$, so $r(u) > r'(u)$. Similarly $r(v) > r'(v)$, and we are done. 
If instead $\XX_dt'$ is not defined on $bu_0$, then $\XX_ap\YY_cq(u) < 
\XX_b(u)$ and $\XX_ap\YY_cq\XX_d(u) = \XX_b\XX_d(u)$, so the condensedness 
of $r = \XX_ap\YY_cq\XX_dt'$ on $u$ implies that $\XX_b\XX_dt'$ is 
condensed on $u$ as well. The same holds for $v$, and we have
\begin{eqnarray*}
\ord(r(u),r'(u)) &=& \ord(\XX_b\XX_dt'(u),r'(u))\textrm{ and similarly,}\\
\ord(r(v),r'(v)) &=& \ord(\XX_b\XX_dt'(v),r'(v)).
\end{eqnarray*}
We conclude by induction on $m$ since $\XX_b\XX_dt'$ has 2 blocks less
than $r$.

\medskip

\noindent\textit{Fourth case: $a \ne b$, $s$ and $s'$ start with
different directions.}\enspace
Without loss of generality, we may assume that $s$ starts with an
$\XX$-block and $s'$ starts with a $\YY$-block.  Since $r$ starts with
2 $\XX$-letters, the number of alternating blocks of $r$ is less than
$|r|-1 \le n-3$.  Therefore if $n = m+1$, $r\in \underR^\XX_{m-2}$ and
$r'\in \underR^\XX_{m-1}$, a case that can be decided by induction on
$m$.  So we now assume that $n \ge m-2$.

If $\XX_b(u) < \XX_a(u)$, then the same inequality holds in $v$ (by
Proposition~\ref{lem:1ranker:1alphabet:samedirection}) and we have
$r'(u) < r(u)$ and $r'(v) < r(v)$.  If instead $\XX_a(u) < \XX_b(u)$
and $\XX_a(v) < \XX_b(v)$, then the $b$-left factorizations of $u_+$
and $v_+$ are of the form $u_+ = u_0bu'_+$ and $v_+ = v_0bv'_+$.

Several cases arise, according to whether $s$ and $s'$ are defined
(and condensed) on $u_0$ or not.  We have $u_+ \RIGHT_{m,n-1} v_+$ and
$u_0 \RIGHT_{m, n-2} v_0$ by Lemma~\ref{cor:leftANDrightNEW}.  It
follows as usual that $s$ and $s'$ are defined on $v_0$ if and only if
they are defined on $u_0$.  If $s$ is not defined on $u_0$ then the
order types $\ord(r(u),r'(u))$ and $\ord(r(v),r'(v))$ are both $>$.
Therefore, from now on we can assume that $s$ is defined on $u_0$ and
$v_0$.

If $s'$ is defined on $u_0$ then we can chop off $u_- a$ from $u$,
$v_- a$ from $v$, and $\XX_a$ from $r$: $\ord(r(u),r'(u)) =
\ord(s(u_+),r'(u_+))$ and $\ord(r(v),r'(v)) = \ord(s(v_+),r'(v_+))$.
Since $u_+ \RIGHT_{m,n-1} v_+$, $\ord(s(u_+),r'(u_+))$ and
$\ord(s(v_+),r'(v_+))$ are equal by induction on $n$, and hence
$\ord(r(u),r'(u)) = \ord(r(v),r'(v))$.

If $s'$ is not defined on $u_0$, then, as in the third case, we have
to split the ranker $s'$ at those points at which it crosses the
position $\XX_a(u)$.  Let $\XX_b s' = p_1 q_1 \cdots p_k q_k$ such
that all $p_i$ are defined on $u_0$ and all $p_i$ are starting with an
$\XX$-letter followed by a (possibly empty) $\YY$-block.  The sole
exception is $p_k$ which might contain further blocks.  Moreover, each
$p_i$ is the maximal prefix of $p_i q_i \cdots p_k q_k$ which is
defined on $u_0$.  All $q_i$ are defined on $u_- a$ and all $q_i$ are
starting with a $\YY$-letter followed by a (possibly empty)
$\XX$-block.  The sole exception is $q_k$ which might be empty or
which might contain further blocks.  Each $q_i$ is the maximal prefix
of $q_i p_{i+1} \cdots p_k q_k$ which is defined on $u_- a$.  Since
$u_- a \RIGHT_{m,n-1} v_- a$ (Lemma~\ref{lem:simplepropd}) and $u_0
\LEFT_{m-1,n-2} v_0$, the same definedness and maximality properties
hold on $v_- a$ and $v_0$.

If $q_k$ is empty, then $k \geq 2$ and $p_1$ and $q_1$ are non-empty.
We see that $\ord(r(u),r'(u)) = \ord(s(u_0),p_k(u_0))$ and
$\ord(r(v),r'(v)) = \ord(s(v_0),p_k(v_0))$.  By induction on $n$, we
have $\ord(s(u_0),p_k(u_0)) = \ord(s(v_0),p_k(v_0))$, and hence
$\ord(r(u),r'(u)) = \ord(r(v),r'(v))$.

Finally, if $q_k$ is non-empty, then we have $r(u) > r'(u)$ and $r(v)
> r'(v)$.
\qed

\subsubsection{Completing the proof of Theorem~\ref{thm: interwoven}}\label{sec: completing the proof}

Let us (at last!) verify that, if $u \RIGHT_{m+1,2n} v$ or $u
\RIGHT_{m+1,2n} v$, then Properties (\textbf{WI 2c}) and (\textbf{WI
3c}) hold for $m,n$.  By symmetry, we simply handle the case where $u
\RIGHT_{m+1,2n} v$.

To verify Property (\textbf{WI 2c}), we consider rankers $r\in
\underR_{m,n}$ and $r' \in \underR_{m-1,n-1}$ that are
condensed on $u$ and $v$.  If both start with $\XX$-blocks,
Proposition~\ref{lem:tworankers:samedirection} shows that
$\ord(r(u),r'(u))$ and $\ord(r(v),r'(v))$ coincide. If both start with 
$\YY$-blocks, the same proposition allows us to conclude, after 
observing that we have $u \LEFT_{m,2n-1} v$. And if $r$ and $r'$ 
start with different direction blocks, we conclude by 
Proposition~\ref{lem:2rankers:differentdirections}.

To verify Property (\textbf{WI 3c}), we consider rankers $r\in
\underR_{m,n}$ and $r' \in \underR_{m,n-1}$ that end with
different directions, and that are condensed on $u$ and $v$.  If $r$
and $r'$ start with different direction blocks, we again conclude by
Proposition~\ref{lem:2rankers:differentdirections}.  If both start
with $\XX$-blocks, then they must have different number of
alternations, so we have $r \in R^\XX_{m_1,n_1}$ and $r' \in
R^\XX_{m_2,n_2}$ for some $n_1 \le n$, $n_2 \le n-1$ and for distinct
values $m_1,m_2 \le m$.  In particular, one of $m_1$ and $m_2$ is less
than or equal to $m-1$, and we can apply
Proposition~\ref{lem:tworankers:samedirection}.

We proceed similarly if $r$ and $r'$ both start with $\YY$-blocks, 
after observing that $u \LEFT_{m,2n-1} v$.
This completes the proof of Theorem~\ref{thm: interwoven}.

\section{Consequences}

\subsection{Decidability results}

The main consequence we draw of Theorem~\ref{thm: interwoven} and of
the decidability of the pseudovarieties $\RR_m$ and $\LL_m$ is 
summarized in the next statement.

\begin{thm}
    Given an $\FO^2$-definable language $L$, one can compute an
    integer $m$ such that $L$ is $\FO^2_{m+1}$-definable, possibly
    $\FO^2_m$-definable, but not $\FO^2_{m-1}$-definable.
    That is: we can decide the quantifier alternation level of $L$
    within one unit.
\end{thm}

\proof
Let $L \in \LangFO^2$ and let $M$ be its syntactic monoid.  Since each
pseudovariety $\RR_m \cap \LL_m$ is decidable (Proposition~\ref{prop from sf}),
 we can compute the largest $m$ such that $M \not\in \RR_m
\cap \LL_m$.  By Theorem~\ref{thm: interwoven}, $M \in \RR_{m+1} \cap
\LL_{m+1} \subseteq \VarFO^2_{m+1}$ and hence $L$ is
$\FO^2_{m+1}$-definable.  On the other hand, $M \not\in \VarFO^2_{m-1}
\subseteq \RR_m \cap \LL_m$.
\qed

Let us also record the following consequences of Proposition~\ref{prop
from sf}, Proposition~\ref{TL2 = R2} and the decidability of $\RR_2 
\vee\LL_2$ (discussed in Example~\ref{example R2L2 vs FO2}).

\begin{prop}
    The classes $\underLangTL^\XX_1 = \underLangTL^\YY_1 =
    \underLangTL_1 = \LangFO^2_1$, $\underLangTL_2^\XX$,
    $\underLangTL_2^\YY$ and $\underLangTL_2$ are decidable.
\end{prop}

\subsection{Infinite and collapsing hierarchies}

The fact that the $\RR_m$ and $\LL_m$ form strict hierarchies
(Proposition~\ref{prop from sf}), together with Theorem~\ref{thm:
interwoven}, proves that the $\LangFO^2_m$ hierarchy is infinite.
Weis and Immerman had already proved this result by combinatorial
means \cite[Theorem 4.11]{WeisImmerman2009lmcs}, whereas our proof is algebraic.
From that result on the $\LangFO^2_m$ hierarchy, it is also possible to recover
the strict hierarchy result on the $\RR_m$ and $\LL_m$ and the fact
that their union is equal to $\DA$.

By the same token, Corollary~\ref{TLm in Rm in FO2m} and
Theorem~\ref{prop: agree 2 condensed} show that the
$\underLangTL_m$ (resp.\  $\underVarTL_m$) hierarchy is infinite and
that its union is all of $\LangFO^2$ (resp.\  $\DA$).

\begin{thm}
    The hierarchies $\LangFO^2_m$ and $\underLangTL_m$ are infinite, 
    and their union is all of $\LangFO^2$.
\end{thm}

Similarly, the fact (stated in Proposition~\ref{prop from sf}) that an
$m$-generated element of $\DA$ lies in $\RR_{m+1}\cap\LL_{m+1}$, shows
that an $\FO^2$-definable language in $A^*$ lies in
$\calR_{|A|+1}\cap\calL_{|A|+1}$, and hence in $\LangFO^2_{|A|+1}$ -- a
fact that was already established by combinatorial means by Weis and
Immerman \cite[Theorem 4.7]{WeisImmerman2009lmcs}.  It also shows that such a
language is in $\underLangTL_{2|A|+1}$ by
Theorem~\ref{prop: agree 2 condensed}.

\begin{thm}\label{fixed A}
    A language $L \subseteq A^*$ is $\FO^2$-definable if and only if
    it is $\FO^2_{|A|+1}$-definable.  And it is $\TL$-definable if and
    only if it is both $\underTL^\XX_{2|A|+1}$ and 
    $\underTL^\YY_{2|A|+1}$-definable.
\end{thm}

Even though we arrived at Theorem~\ref{fixed A} by algebraic
means, it is interesting to note that its statement reflects the
following combinatorial property (an idea that was already used by
Weis and Immerman \cite[Theorem 4.7]{WeisImmerman2009lmcs}).

\begin{lem}\label{fixed alphabet}
    A ranker that is condensed on a word on alphabet $A$, has at most
    $|A|$ alternating blocks.
\end{lem}

\proof
Let $u$ be a word and let $r$ be a ranker that is 
condensed on $u$. Without loss of generality, we may assume that $r 
\in R_{m,n}^\XX$, say
$$r = \XX_{a_1}\cdots\XX_{a_{k_1}} \YY_{a_{k_1+1}}\cdots 
\YY_{a_{k_2}} \cdots \ZZ_{a_{k_{m-1}+1}}\cdots \ZZ_{a_{k_m}}$$
with $0 < k_1 < k_2 < \cdots < k_m = n$ and $\ZZ = \XX$ (resp.\  $\YY$)
if $m$ is odd (resp.\  even).  By definition of condensed rankers (and
with the notation in that definition, see Section~\ref{sec:
condensed rankers}), the interval $I_{k_h}$ is of the form
$(i_{k_h-1}; \XX_{a_{k_h}}(u,i_{k_h-1}))$
if $h$ is odd, of the form
$(\YY_{a_{k_h}}(u,j_{k_h-1}); j_{k_h-1})$
 if $h$ is even.  In either
case, $a_{k_{h+1}}$ occurs in $u$ within the interval $I_{k_{h}}$ but 
$a_{k_h}$ does not.  Since the intervals $I_{k_h}$ are
nested, it follows that the letters $a_{k_1}, a_{k_2},\ldots, a_{k_m}$
are pairwise distinct, and hence $m \le |A|$.
\qed

\subsection{Infinite hierarchies and unambiguous polynomials}\label{sec: unambiguous}

Finally we note the following refinement of \cite[Proposition~4.6]{KufleitnerWeilSF2010}.  One of the classical (and one of 
the earliest) results concerning the languages recognized by 
monoids in $\DA$ is the following: they are 
exactly the disjoint unions of 
unambiguous products of the form
$B_0^*a_1B_1^*\cdots a_kB_k^*$, where each $B_i$ is a subset of $A$ 
(Sch\"utzenberger \cite{Schutzenberger1976sf}, see also 
\cite{TessonTherien2002,TessonTherien2007lmcs,DiekertGK2008ijfcs}). Recall that such a product is \emph{unambiguous} if each word $w\in B_0^*a_1B_1^*\cdots a_kB_k^*$ factors in a unique way as $w = u_0a_1u_1\cdots a_ku_k$ with $u_i\in B_i^*$. Deterministic and co-deterministic products 
(see Section~\ref{sec: deterministic}) are easily seen to be 
particular cases of unambiguous products.
Propositions~\ref{malcev language statement} and~\ref{prop from sf} 
imply the following statement.

\begin{prop}
    The least variety of languages containing the languages of the 
    form $B^*$ ($B \subseteq A$) and closed under visibly 
    deterministic and visibly co-deterministic products, is 
    $\LangFO^2$.
    
    More precisely, every unambiguous product of the form
    $B_0^*a_1B_1^*\cdots a_kB_k^*$, where each $B_i$ is a subset of
    $A$, can be expressed in terms of 
    Boolean operations and at most $|A|+1$ alternated applications of
    visibly deterministic and visibly co-deterministic products -- 
    starting with a visibly deterministic (resp.\ co-deterministic) 
    product.
\end{prop}

The analogous, but weaker statement with the word \textit{visibly}
deleted was proved in \cite{KufleitnerWeilSF2010} by algebraic 
means, and independently by Lodaya, Pandya and Shah using logical and 
combinatorial arguments \cite{LodayaPS2008ifip}.

\subsection*{Conclusion}

We have related the $\FO^2_m$ hierarchy with the $\calR_m$-$\calL_m$ hierarchy, a hierarchy of 
varieties of languages which is connected with the alternation of closures under deterministic and co-deterministic products.

The varieties $\calR_m$ and $\calL_m$ are decidable, but the link we establish with $\LangFO^2_m$ (Theorem~\ref{thm: interwoven}) is not tight enough to prove decidability of the quantifier alternation hierarchy. We recall the readers of our conjecture (Conjecture~\ref{conjecture FO2} above), according to which
$\LangFO^2_m$ is equal to the intersection $\calR_{m+1} \cap \calL_{m+1}$.
Establishing this conjecture would prove that each level of the quantifier alternation hierarchy $\LangFO^2_m$ is decidable.

Finally, we refer the reader to Straubing's result: he showed \cite{2011:Straubing} that the pseudovariety $\VarFO^2_m$ is the $m$-th weakly iterated power of the pseudovariety $\J$ of $\calJ$-trivial monoids (more precisely, $\VarFO^2_1 = \J$ and $\VarFO^2_{m+1} = \VarFO^2_m \block \J$). This result offers a different avenue to solve the decidability problem for $\FO^2_m$-definability, and our conjecture would show the equality between two algebraic hierarchies which seem completely unrelated.

\section*{Acknowledgements}

The authors gratefully acknowledge the contribution of the referees, which helped clarify certain points in the paper, suggested simpler proofs for certain technical lemmas and --- especially! --- pointed out a mistake in one of the proofs.

%
%

\end{document}